\documentclass[useAMS,usenatbib]{mn2e}
\usepackage{amssymb,amsmath,amsfonts,graphicx}
\usepackage{epsfig,color} 
\usepackage{xcolor,colortbl}
\usepackage[section] {placeins}
\usepackage[T1]{fontenc}
\usepackage{aecompl}
\usepackage{times}
\usepackage{mathtools}

\newcommand{\ud}{\mathrm{d}}
\newcommand{\rc}{\rowcolor{lightgray}}
\newcommand\avg[1]{{\langle #1 \rangle}}

\def\lsim{ \lower .75ex \hbox{$\sim$} \llap{\raise .27ex \hbox{$<$}} }

\title[]{Estimating the dark matter halo mass of our Milky Way using
  dynamical tracers}

\author[Wang et al.]{Wenting Wang$^{1}$, Jiaxin Han$^{1}$ , Andrew P. Cooper$^{1}$, Shaun Cole$^{1}$, Carlos Frenk$^{1}$, \and
Ben Lowing$^{1}$ \\
  {}$^{1}$Institute for Computational Cosmology, University of Durham, South Road, Durham, DH1 3LE, UK
}

\begin{document}



\maketitle

\begin{abstract}
  The mass of the dark matter halo of the Milky Way can be estimated
  by fitting analytical models to the phase-space distribution of
  dynamical tracers. We test this approach using realistic mock
  stellar halos constructed from the Aquarius N-body simulations of
  dark matter halos in the $\Lambda$CDM cosmology. We extend the
  standard treatment to include a Navarro-Frenk-White (NFW) potential
  and use a maximum likelihood method to recover the parameters
  describing the simulated halos from the positions and velocities of
  their mock halo stars. We find that the estimate of halo mass is
  highly correlated with the estimate of halo concentration. The best-fit
  halo masses within the virial radius, $R_{200}$, are biased, ranging
  from a 40\% underestimate to a 5\% overestimate in the best case
  (when the tangential velocities of the tracers are included). There 
  are several sources of bias. Deviations from dynamical equilibrium 
  can potentially cause significant bias; deviations from spherical 
  symmetry are relatively less important. Fits to stars at different 
  galactocentric radii can give different mass estimates. By contrast, 
  the model gives good constraints on the mass within the half-mass 
  radius of tracers even when restricted to tracers within 60~kpc. 
  The recovered velocity anisotropies of tracers, $\beta$, are biased 
  systematically, but this does not affect other parameters if tangential 
  velocity data are used as constraints.
\end{abstract}

\begin{keywords}
Galaxy: Milky-Way
\end{keywords}

\section{Introduction}
\label{sec:intro}

Our Milky Way (MW) galaxy provides a wealth of information on the
physics of galaxy formation and the nature of the dark matter. This
information can, in principle, be unlocked from studies of the
positions, velocities and chemistry of stars in the Galaxy, its
satellites and globular clusters, which can be observed with high
precision.

Many inferences derived from the properties of the Milky Way (MW)
depend on the precision and accuracy with which the mass of its dark
matter halo can be estimated. An example is the much-publicised ``too
big to fail'' problem, the apparent lack of MW satellite galaxies with
central densities as high as those of the most massive dark matter
subhalos predicted by $\mathrm{\Lambda}$CDM simulations of `Milky Way
mass' hosts \citep{2011MNRAS.415L..40B,2012MNRAS.422.1203B,2012MNRAS.425.2817F}. 
In these simulations the number of massive subhalos depends strongly on
the assumed MW halo mass and the problem disappears if the MW halo
mass is sufficiently small \citep[$\lsim 1 \times
10^{12}$~M$_\odot$;][]{2012MNRAS.424.2715W, Cautun_2014a}.

Gravitational lensing is the most powerful method to determine the 
underlying dark matter distribution for large samples of distant galaxies 
\citep[e.g.][]{2001PhR...340..291B,Mandelbaum2006a,2009MNRAS.394.1016L,
2010MNRAS.404..486H,2015MNRAS.446.1356H}. Our MW is, however, special because 
we are embedded in it, and there are many different ways of constraining the 
MW dark matter halo mass\footnote{We use $M_{200}$ and $R_{200}$ to denote 
the mass and radius of a spherical region with mean density equal to 200 
times the critical density of the Universe. }.

These methods include timing argument estimators \citep{1959ApJ...130..705K} 
calibrated against $N$-body simulations \citep{2008MNRAS.384.1459L}; modeling 
of local cosmic expansion \citep{2014MNRAS.443.2204P}; the kinematics of bright 
satellites \citep{2007MNRAS.379.1464S,2007MNRAS.379.1475S,2014MNRAS.437..959B,
Cautun_2014b}, particularly Leo I \citep{2013ApJ...768..140B} and the Magellanic 
Clouds \citep{2011ApJ...743...40B, 2013ApJ...770...96G}; the kinematics of 
stellar streams \citep{2010ApJ...711...32N,2015ApJ...803...80K}, especially the 
Sagittarius stream \citep{2005ApJ...619..807L,2014MNRAS.445.3788G}; measurements of 
the escape velocity using nearby high velocity stars, such as those from the RAVE 
survey \citep{2007MNRAS.379..755S, 2014A&A...562A..91P}; and combinations of
photometric and kinematic data such as Maser observations and terminal velocity
curves \citep{2011MNRAS.414.2446M,2013JCAP...07..016N}. Using high 
resolution hydrodynamical simulations and the line-of-sight velocity dispersion 
of tracers in the MW, \cite{2013ApJ...773L..32R} found a heavy MW halo mass 
reported in some previous measurements of $M_{200} \approx 2\times10^{12}\mathrm{M_\odot}$ 
is disfavoured. 

Some authors have used large composite samples of objects assumed to
be dynamical tracers in the halo, such as stars, globular clusters and
planetary nebulae. For example, the halo circular velocity,
$V_\mathrm{circ}$, may be inferred from the radial velocity dispersion
of tracers, $\sigma_r(r)$, using the spherical Jeans equation. Such
methods require the tracer velocity anisotropy and density profiles to
be known or assumed.  \cite{2005MNRAS.364..433B} made use of a few
hundred stars and globular clusters from 20 to 120~kpc;
\cite{2008ApJ...684.1143X} used 2401 BHB stars from SDSS/DR6 ranging
from 20 to 60~kpc; \cite{2010ApJ...720L.108G} used BHB and RR Lyrae
stars ranging from 25 to 80~kpc; and \cite{2010MNRAS.406..264W} used
26 satellites within 300~kpc with tracer mass estimators, with the method 
further improved by \cite{2011ApJ...730L..26E} and \cite{2012MNRAS.420.2562A}. 
Most recently \cite{2012ApJ...761...98K,2014ApJ...794...59K} used a few 
thousand BHB stars extending to 60~kpc and K-giants beyond 100~kpc.

Most measurements based on dynamical tracers involve assumptions about
the tracer density profiles and velocity anisotropies. However,
\cite{1999MNRAS.310..645W} introduced a Bayesian likelihood analysis,
based on fitting a model phase-space distribution function to the
observed distances and velocities of tracers. In their analysis the
tracer density profile and velocity anisotropy can be considered as
free parameters of the distribution function, to be constrained
together with parameters of the host halo such as its mass and
characteristic scalelength. The sample of stars used by
\cite{1999MNRAS.310..645W} was small and their best-fit host halo mass
for a truncated flat rotation curve model was $1.9_{-1.7}^{+3.6}
\times 10^{12}$~M$_\odot$ (see also \citealt{2003A&A...397..899S}). 
More recently, \cite{2012MNRAS.424L..44D} used a few thousand BHB 
stars from SDSS up to $r\sim$50~kpc. \cite{2015ApJ...806...54E} 
introduced a generalised Bayesian approach to deal with incomplete 
data, which avoids rewriting the distribution function when tangential 
velocities are not available.


\begin{figure} \epsfig{figure=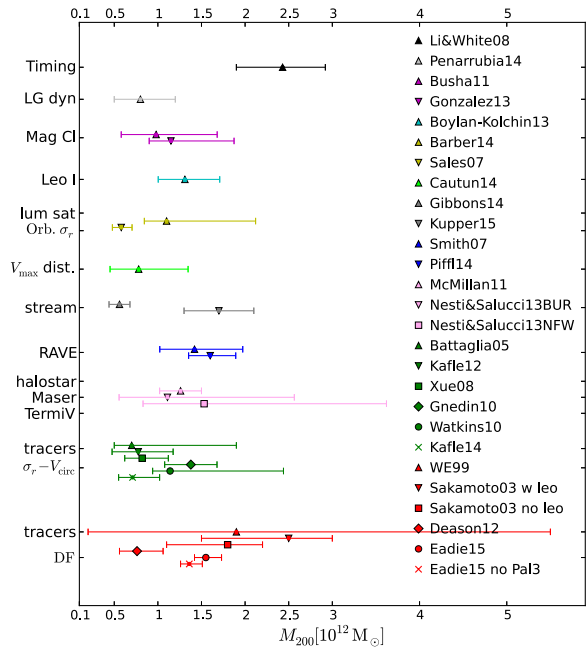,width=0.49\textwidth}%
  \caption{Measured MW halo masses in the literature ($x$-axis), converted
  to $M_{200}$ and categorised by methodology ($y$-axis). Measurements using
  similar methods and/or tracer populations are plotted with the same colour.
  Categories include the timing argument estimator (black); a model of the 
  local cosmic expansion (light grey); constraints from
  luminous MW satellites such as the Magellanic Clouds (magenta), Leo I(cyan),
  the orbits or radial velocity dispersion of other bright satellites (yellow)
  and their $V_\mathrm{max}$ distribution (light green); modelling of 
  tidal streams (grey); high velocity stars from the RAVE survey
  (blue); combinations of maser observations and and the terminal velocity
  curve (pink); and (of most relevance to this work) dynamical modelling using
  large samples of dynamical tracers (red and green).  Methods involving large 
  samples of dynamical tracers are split into two categories, 1) those based on 
  the radial velocity dispersion of tracers (green) and 2) those using model 
  distribution function to constrain both halo properties and velocity anisotropies 
  of tracers simultaneously (red). We have converted results to $M_{200}$ by
  assuming an NFW density profile and a common mass-concentration relation.
  95\% or 90\% confidence intervals have been converted to 1$\sigma$ errors by
  assuming a Gaussian error distribution, except for Watkins et al. and 
  measurements in the DF classification. 
   }
 \label{fig:mass} 
\end{figure}

Fig.~\ref{fig:mass} summarises the results of these studies. The
$x$-axis is the measured MW halo mass. We have converted results to
$M_{200}$ by assuming an NFW density profile
\citep{Navarro_1996,Navarro_1997} and using the mean halo
concentration relation of \cite{2008MNRAS.390L..64D} in cases where a
value for concentration is not given in the original study. The
measurements are grouped by methodology, indicated by colours and
labeled along the $y$-axis. We group those methods that use large samples 
of dynamical tracers into two sets: 1) those based on the radial velocity
dispersion of the tracers and spherical Jeans equation to infer the 
circular velocity and underlying potential; 2) those based on fitting to 
model distribution functions, which attempt to constrain both halo mass 
and the velocity anisotropy of the tracers simultaneously.  Errorbars
correspond to those quoted by the original authors; we have converted
90\% or 95\% confidence intervals to 1$\sigma$ errors assuming a
Gaussian distribution, except for Watkins et al. and measurements in 
the DF classification. This is because \cite{1999MNRAS.310..645W},
\cite{2003A&A...397..899S}, \cite{2010MNRAS.406..264W},  
and \cite{2012MNRAS.425.2840D} included
other sources of model uncertainties beyond pure statistical errors in 
their measured masses, which makes their errors relatively large. For 
the generalised Bayesian approach of \cite{2015ApJ...806...54E}, we 
quote the 95\% Bayesian confidence interval. Fig.~\ref{fig:mass} shows 
that existing measurements of the most likely MW halo mass differ by 
more than a factor of 2.5, even when similar methods are used, although 
apart from a few outliers, the estimates are statistically consistent.

Here we are particularly interested in methods such as that of
\cite{1999MNRAS.310..645W}, which treat the spatial and dynamical
properties of tracers as free parameters to be constrained under the
assumption of theoretical phase-space distribution functions. The
primary aim of this paper is to test the model distribution functions
used in this approach. We extend the distribution function proposed by
\cite{1999MNRAS.310..645W} to one based on the NFW potential, and
model the radial profiles of tracers with a more general double
power-law functional form.  The model function is then fit to the
phase-space distribution of stars in realistic mock stellar halo
catalogues constructed from the cosmological galactic halo simulations
of the Aquarius project \citep{Springel_2008}, to understand its
reliability and possible violations to the underlying assumptions. Our
results have implications that are not limited to the specific form of
the distribution function that we test, but are applicable to the
method itself. 

This paper is structured as follows. The mock stellar halo catalogues
are introduced in Section~\ref{sec:mock}.  Detailed descriptions of
the model distribution function and the maximum likelihood approach
are provided in Section~\ref{sec:method}. Our results are presented in
Section~\ref{sec:6para}, with detailed discussions of reliability and
systematics in Section~\ref{sec:sourcesofbias} and
Section~\ref{sec:rbin}. We conclude in
Section~\ref{sec:concl}. Throughout this paper we adopt the cosmology
of the Aquarius simulation series ($H_0=73
~\mathrm{km~s^{-1}}~\mathrm{Mpc}^{-1}$, $\Omega_{\rm m}=0.25$,
$\Omega_\Lambda=0.75$ and $n=1$).

\section{Mock stellar halo catalogue}
\label{sec:mock}

We use mock stellar halo catalogues constructed from the Aquarius
N-body simulation suite \citep{Springel_2008} with the particle
tagging method described by \citet{2010MNRAS.406..744C}, to which we refer
the reader for further details.  In this section we summarise the most
important features of these catalogues.

\subsection{The Aquarius simulations}

The Aquarius halos come from dark matter N-body simulations in a
standard $\mathrm{\Lambda}$CDM cosmology. Cosmological parameters are
those from the first year data of WMAP \citep{2003ApJS..148..175S}.
Our work uses the second highest resolution level of the Aquarius
suite, which corresponds to a particle mass of $\sim10^4
h^{-1}$~M$_\odot$.

The simulation suite includes six dark matter halos with virial
masses spanning the factor-of-two range of Milky Way observations
discussed in the previous section.  We have only used five out of the
six halos for our analysis (labeled halo A to halo E according to
the Aquarius convention). The halo we have not used (halo F) undergoes
two major merger events at $z<0.6$, and is thus an unlikely host for a
MW-like disc galaxy. We list in Table~\ref{tbl:parameters} the host
halo mass, $M_{200}$, and other properties of the five halos, which
are taken from \cite{2010MNRAS.402...21N}.

\subsection{The galaxy formation and evolution model}

The Durham semi-analytical galaxy formation model, GALFORM, has been
used to post-process the Aquarius simulations, predicting the
evolution of galaxies embedded in dark matter halos. To construct the
mock stellar halo catalogues used in this paper, the version described
by \cite{2011MNRAS.417.1260F} was adopted.  This model has several
minor differences from the model of \cite{2006MNRAS.370..645B}, such
that the \cite{2011MNRAS.417.1260F} model matches better the observed
luminosity function, luminosity-metallicity relation and radial
distribution of MW satellites. The main changes are a more
self-consistent calculation of the effects of the patronisation
background and a higher chemical yield in supernovae feedback.

\subsection{Particle tagging}
\label{sec:tagging}

The GALFORM model predicts the amount of stellar mass present in each 
dark matter halo in the simulation at each output time, as well as properties of
stellar populations such as their total metallicity.  However, GALFORM does not
provide detailed information about how these stars are distributed in galaxies.
The particle tagging method of \cite{2010MNRAS.406..744C} is a way to determine
the six-dimensional spatial and velocity distribution of stars from dark matter
only simulations, by associating newly-formed stars with tightly bound dark
matter particles.

At each simulation snapshot, each newly formed stellar population
predicted by GALFORM is assigned to the 1\% most bound dark matter
particles in its host dark matter halo. Each ``tagged'' dark matter
particle then represents a fraction of a single stellar population,
the age and metallicity of which are also known from GALFORM. Traced
forward to the present day, these tagged particles give predictions
for the observed luminosity functions and structural properties of MW
and M31 satellites that match well to observations.  Recently,
\cite{2013MNRAS.434.3348C} have applied this technique to large-scale
cosmological simulations and have shown that it produces galactic
surface brightness profiles that agree well with the outer regions of
stacked galaxy profiles from SDSS.

Our study is based on tagged dark matter particles from accreted
satellite galaxies. We ignore particles associated with in situ star
formation in the central galaxy. Strictly, our results thus only apply
in the case where most MW halo stars originate from accretion. This is
supported by the data of \cite{2008ApJ...680..295B,2010AJ....140.1850B}
although other work suggests that a certain fraction of the halo stars
are contributed by in-situ star formation, especially close to the
central galaxy ($r<30$~kpc) \citep[see,
e.g.][]{2007Natur.450.1020C,2010ApJ...712..692C,2010ApJ...721..738Z,2011ApJ...733L...7H}.
Ignoring the possible in-situ component is thus a weakness of our mock
stellar halo catalogue. Nevertheless, our mock halo stars enable us to
test and constrain the theoretical distribution function and, in
practice, most of our conclusions (see Sec.~\ref{sec:6para} and
Sec.~\ref{sec:sourcesofbias}) do not depend on whether the MW halo
stars formed in-situ or were brought in by accretion.

\begin{figure}
\epsfig{figure=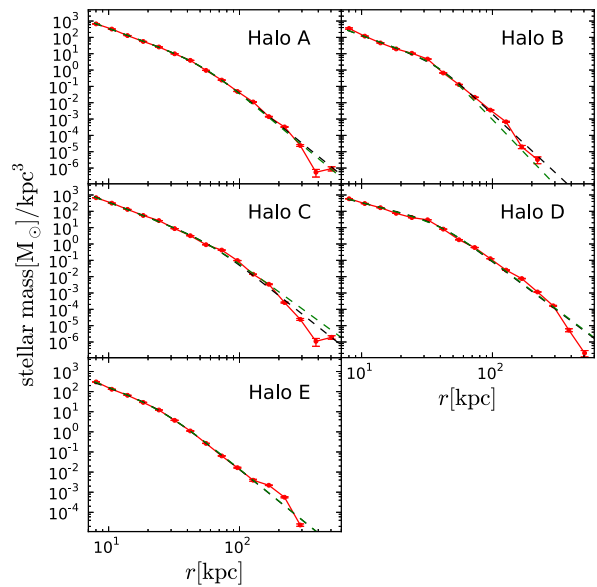,width=0.49\textwidth}%
\caption{The radial density profiles of stellar mass (red points and
  lines with errors) in the mock stellar halo catalogue of the five
  Aquarius halos. The errors are in most cases of comparable size to
  the symbols and are almost invisible. Dashed black curves are double
  power-law fits to the red data points obtained from a $\chi^2$
  minimisation. Green dashed curves are the best-fit density profiles
  from the maximum likelihood method.}
\label{fig:densp}
\end{figure}

\section{Methodology}
\label{sec:method}

In this section we discuss the theoretical context of our method for
constraining dark matter halo properties using dynamical tracers and a maximum
likelihood approach based on theoretical distribution functions. In
Sec.~\ref{sec:theory}, we describe how the phase-space distribution of the
tracer population is modeled. Sec.~\ref{sec:nfw} gives details about the
explicit form of the distribution function. The likelihood function is
introduced in Sec.~\ref{sec:like}. Finally, we describe how we weight tagged
particles and how errors are estimated in Sec.~\ref{sec:weight}.  Our method
follows that of \cite{1999MNRAS.310..645W} but introduces significant
modifications to the form of the dark matter halo potential and the assumed
tracer density profile. 

\subsection{Phase-space distribution of Milky Way halo stars}
\label{sec:theory}

The phase-space distribution function of tracers (e.g. stars) bound to a dark
matter halo potential (binding energy $E>0$) can be described by the Eddington
formula \citep{1916MNRAS..76..572E}. The simplest isotropic and spherically
symmetric case is 

\begin{equation}
F(E)=\frac{1}{\sqrt{8}\pi^2}\frac{\ud}{\ud E}\int_{\Phi(r_\mathrm{max,t})}^E \frac{\ud \rho(r)}{\ud\Phi(r)}\frac{\ud\Phi(r)}{\sqrt{E-\Phi(r)}},
\label{eq:isoform} 
\end{equation} where the distribution function only depends on the binding energy per unit
mass, $E=\Phi(r)-\frac{v^2}{2}$. $\Phi(r)$ and ${v^2}/{2}$ are the
underlying dark matter halo potential and kinetic energy per unit mass of
tracers. The integral goes from the potential at the tracer boundary\footnote{To
define the binding energy, we adopt the convention that $\Phi(r)>0$.} to the 
binding energy of interest.  Usually both the zero point of potential and tracer 
boundary, $r_\mathrm{max,t}$, are chosen at infinity, and thus
$\Phi(r_\mathrm{max,t})=0$. 

In reality the velocity distribution of tracers may be anisotropic
and depend both on energy  and angular
momentum, $L$. In the simplest case, the distribution function is
assumed to be separable: 

\begin{equation}
F(E,L)=L^{-2\beta}f(E),  
\label{eq:anisoform}
\end{equation}
where the energy part, $f(E)$, is expressed as \citep{1991MNRAS.253..414C}

\begin{equation}
\begin{split}
f(E)=\frac{2^{\beta-3/2}}{\pi^{3/2}\Gamma(m-1/2+\beta)\Gamma(1-\beta)}\times\\
\frac{\ud}{\ud E}\int_{\Phi(r_\mathrm{max,t})}^E (E-\Phi)^{\beta-3/2+m} \frac{\ud^m [r^{2\beta}\rho(r)]}{\ud \Phi^m} \ud \Phi\\
=\frac{2^{\beta-3/2}}{\pi^{3/2}\Gamma(m-1/2+\beta)\Gamma(1-\beta)}\times\\
\int_{\Phi(r_\mathrm{max,t})}^E (E-\Phi)^{\beta-3/2+m} \frac{\ud^{m+1} [r^{2\beta}\rho(r)]}{\ud \Phi^{m+1}} \ud \Phi.
\end{split}
\end{equation} Here $\beta$ is the velocity anisotropy parameter defined as 

\begin{equation}
\beta=1-\frac{\avg{{v_\theta}}^2-\avg{v_\theta}^2+\avg{{v_\phi}^2}-
\avg{v_\phi}^2}{2(\avg{{v_r}^2}-\avg{v_r}^2)}, 
\label{eqn:beta}
\end{equation}  with $v_r$, $v_\theta$ and $v_\phi$ being the radial and two tangential components of the velocity. The integer, $m$, is chosen 
to make the integral converge and depends on the value
of $\beta$. In our analysis the parameter range of $\beta$ is $-0.5<\beta<1$
and $m=1$. $\beta>0$ represents radial orbits, while tangential
orbits have $\beta<0$. $\beta=0$ corresponds to the isotropic velocity
distribution.

In real observations, the tangential velocities of tracers are often
unavailable. We thus test two different cases, in which i) only radial
velocities are available and ii) both radial and tangential velocities are
available. For case (i), the phase-space distribution in terms of radius, $r$,
and radial velocity, $v_r$, is given by the integral over tangential velocity,
$v_t=\sqrt{v_\theta^2+v_\phi^2}$, as

\begin{equation}
P(r,v_r|C)=\int L^{-2\beta}f(E) 2\pi v_t \ud v_t,
\end{equation}
where $C$ denotes a set of model parameters. With the Laplace transform, this can be written as

\begin{equation}
P(r,v_r|C)=\frac{1}{\sqrt{2}\pi r^{2\beta}} \int_{\Phi(r_\mathrm{max,t})}^{E_r} \frac{\ud\Phi}{\sqrt{E_r-\Phi}}\frac{\ud r^{2\beta}\rho(r)}{\ud\Phi},
\label{eq:radialonly}
\end{equation}
where $E_r=\Phi(r)-v_r^2/2$. All factors of $m$ cancel in the Laplace transform
and hence Eqn.~\ref{eq:radialonly} does not depend on $m$. For case (ii), the
distribution function is simply Eqn.~\ref{eq:anisoform}, i.e.

\begin{equation}
P(r,v_r,v_t|C)=L^{-2\beta}f(E), 
\label{eq:RT} 
\end{equation}
where $E=\Phi(r)-v_r^2/2-v_t^2/2$ and $L=r v_t$.

\subsection{NFW potential and double power-law density profiles of the tracer population}
\label{sec:nfw}

\cite{1999MNRAS.310..645W} and \cite{2003A&A...397..899S} adopted the so-called
truncated flat rotation curve model for the underlying dark matter potential.
In our analysis, we will extend Eqn.~\ref{eq:anisoform} to the NFW potential
\citep{Navarro1996,Navarro1997}

\begin{equation}
\Phi(r)=-4\pi G \rho_s r_s^2 \left(
\frac{\ln(1+r/r_s)}{r/r_s}+\frac{1}{1+r_\mathrm{max,h}/r_s} \right),
\label{eq:potential}
\end{equation}
when $r<r_\mathrm{max,h}$, and 
\begin{equation}
\Phi(r)=-4\pi G \rho_s r_s^2 \left(
\frac{\ln(1+r_\mathrm{max,h}/r_s)}{r/r_s}+\frac{r_\mathrm{max,h}/r_s}{(r/r_s)(1+r_\mathrm{max,h}/r_s)}\right),
\label{eq:potential2}
\end{equation}
when $r>r_\mathrm{max,h}$.

There are two parameters in Eqn.~\ref{eq:potential} and
Eqn.~\ref{eq:potential2}, the scalelength, $r_s$, and the scaledensity, $\rho_s$,
defined at $r=r_s$. $r_\mathrm{max,h}$ is the halo boundary. If the halo is
infinite, the second term in Eqn.~\ref{eq:potential} vanishes. In most
of our analysis, we will assume the NFW halo is infinite. We test different
choices of halo boundary in the Appendix~\ref{sec:fini}.  

To derive analytical expressions for Eqn.~\ref{eq:radialonly} and
Eqn.~\ref{eq:RT}, we need an analytical form for the tracer density profile,
$\rho(r)$. Fig.~\ref{fig:densp} shows the radial density profile of stellar
mass (red points) in each of the five Aquarius halos. Error bars are obtained
from 100 realisations of bootstrap resampling. In most of the cases, these
profiles can be described well by a double power law (black dashed lines are
double power-law fits that minimise $\chi^2$). Significant deviations from a
double power law are most obvious in the outskirts of the halos. For example,
halo E has a prominent bump at $r\sim100$~kpc due to a tidal stream.

There are indications that the real MW has a two-component profile, with
density falling off more rapidly beyond $\sim25$~kpc, whereas M31 has a smooth
profile out to 100~kpc with no obvious break
\citep[e.g.][]{2009MNRAS.398.1757W, 2011MNRAS.416.2903D, 2011ApJ...731....4S}.
Recently, \cite{2014ApJ...787...30D} report evidence for a very steep outer
halo profile of the MW. If we believe that MW halo stars originate from the
accretion of dwarf satellites, whether the profile is broken or unbroken
depends on the details of accretion history \citep{2013ApJ...763..113D,
2015MNRAS.446.2274L}. There is an as yet unresolved debate over whether the
stellar halo of the MW has an additional contribution from stars formed in
situ, in which case a break in the profile may reflect the transition from in
situ-dominated regions to accretion dominated regions. 

As our mock halo stars (which are all accreted) and observed MW halo stars can
be approximated by a double power-law profile, we adopt the following
functional form to model tracer density profiles: 

\begin{equation}
\rho(r)\propto \left[\left(\frac{r}{r_0}\right)^\alpha+\left(\frac{r}{r_0}\right)^\gamma\right]^{-1}.
\end{equation}
This equation has three parameters: the inner slope, $\alpha$, the outer slope,
$\gamma$, and the transition radius, $r_0$.  

Previous studies have adopted a single power law to describe the density
profile of MW halo stars beyond $r\sim20$~kpc
\citep[e.g.][]{2008ApJ...684.1143X,2010ApJ...720L.108G, 2012MNRAS.424L..44D,
1999MNRAS.310..645W}. Our double power-law form naturally includes this
possibility as a special case. We also note that \cite{2003A&A...397..899S}
considered the case of ``shadow'' tracers with a radial distribution that
shares the same functional form with the underlying dark matter. We emphasise 
that our mock halo stars are not ``shadow'' tracers; their radial distribution 
is significantly different from that of the dark matter.

Assuming these analytical expressions for $\Phi(r)$ and $\rho(r)$,
Eqn.~\ref{eq:radialonly} and Eqn.~\ref{eq:RT} can be written more explicitly as

\begin{equation}
\begin{split}
 P(r,v_r|\rho_s,r_s,\beta,\alpha,\gamma,r_0)=-\frac{r_s^{2\beta-\alpha-\gamma}}{\sqrt{2}\pi r^{2\beta} v_s}\int_{R_{\mathrm{inner}}}^{R_\mathrm{max,t}} \frac{R'^{2\beta-1}}{\sqrt{\epsilon(r)-\phi(R')}}\\ 
 \times \frac{(2\beta-\alpha)(\frac{R'}{r_0})^\alpha r_s^{-\gamma}+(2\beta-\gamma)(\frac{R'}{r_0})^\gamma r_s^{-\alpha}}{[(\frac{R'}{r_0})^\alpha r_s^{-\gamma}+(\frac{R'}{r_0})^\gamma  r_s^{-\alpha}]^2} \ud R',
\end{split}
\label{eq:radialonly2}
\end{equation}
and
\begin{equation}
\begin{multlined}
 P(r,v_r,v_t|\rho_s,r_s,\beta,\alpha,\gamma,r_0)=\\
-\frac{r_s^{-\alpha-\gamma}l^{-2\beta}}{2^{3/2-\beta}\pi^{3/2} v_s^3\Gamma(\beta+1/2)\Gamma(1-\beta)} \times \\
 \int_{R_{\mathrm{inner}}}^{R_\mathrm{max,t}} \ud R' (\epsilon(r)-\phi(R'))^{\beta-1/2} \times \\
\Bigg\{ \frac{(2\beta+1) R'^{2\beta}\left(\frac{R'}{1+R'}-\ln(1+R')\right)-\left[\frac{1}{(1+R')^2}-\frac{1}{1+R'}\right]R'^{2\beta+1}}{\left[\frac{R'}{1+R'}-\ln(1+R')\right]^2}\\
\times \frac{(2\beta-\alpha)\left(\frac{R'}{r_0}\right)^\alpha
  r_s^{-\gamma}+(2\beta-\gamma)\left(\frac{R'}{r_0}\right)^\gamma \,
  r_s^{-\alpha}}{\left[\left(\frac{R'}{r_0}\right)^\alpha \,
    r_s^{-\gamma}+\left(\frac{R'}{r_0}\right)^\gamma  \, r_s^{-\alpha}\right]^2}+ \\
\frac{R'^{2\beta+1}}{\left[\frac{R'}{1+R'}-\ln(1+R')\right]\left[\left(\frac{R'}{r_0}\right)^\alpha
    \, r_s^{-\gamma}+\left(\frac{R'}{r_0}\right)^\gamma \,  r_s^{-\alpha}\right]^3}\times \\
\Bigg[(2\beta-\alpha)r_s^{-\alpha-\gamma}\left(\frac{\alpha}{r_0}-\frac{2\gamma}{r_0}\right)\left(\frac{R'}{r_0}\right)^{\alpha+\gamma-1}+ \\
(2\beta-\gamma)r_s^{-\alpha-\gamma}\left(\frac{\gamma}{r_0}-\frac{2\alpha}{r_0}\right)\left(\frac{R'}{r_0}\right)^{\alpha+\gamma-1}- \\
(2\beta-\alpha)r_s^{-2\gamma}\frac{\alpha}{r_0}\left(\frac{R'}{r_0}\right)^{2\alpha-1}-(2\beta-\gamma)r_s^{-2\alpha}\frac{\gamma}{r_0}\left(\frac{R'}{r_0}\right)^{2\gamma-1}
\Bigg]
\Bigg\}.
\end{multlined}
\label{eq:RT2}
\end{equation}
Here,
analogously to \cite{1999MNRAS.310..645W}, we have introduced a characteristic
velocity, $v_s=r_s\sqrt{4\pi G \rho_s}$. The binding energy, $\epsilon$,
angular momentum, $l$, potential, $\phi$, and radius, $R$, have all been scaled
by $v_s$ and $r_s$ and are thus dimensionless, as follows, 
\begin{eqnarray}
 \epsilon=\frac{E}{v_s^2},\ l=\frac{L}{r_s v_s},\ \phi=\frac{\Phi}{v_s^2},\ R=\frac{r}{r_s}.
\label{eq:dimless}
\end{eqnarray}
As mentioned above, $R_\mathrm{max,t}$ is the boundary of the tracer
distribution and, for most of
our analysis, we take $R_\mathrm{max,t}=\infty$.  Note that
Eqn.~\ref{eq:radialonly2} and Eqn.~\ref{eq:RT2} are deduced by assuming the
tracer boundary, $R_\mathrm{max,t}$, is smaller or equal to the halo boundary,
$R_\mathrm{max,h}$. In both Eqn.~\ref{eq:radialonly2} and Eqn.~\ref{eq:RT2}
there are six model parameters. 

The phase-space probability of a tracer at radius, $r$, whose radial and
tangential velocities are $v_r$ and $v_t$, can be derived from
Eqn.~\ref{eq:radialonly2} or Eqn~\ref{eq:RT2}. The lower limit of the integral
is determined by solving
\begin{equation} 
\phi(R_{\mathrm{inner}})=\epsilon, 
\end{equation} 
where $\epsilon$ equals $\phi(R)-{v_r^2}/{(2v_s^2)}$ when only the radial velocity
  is available, and $\epsilon$ equals 
  $\phi(R)-{v_r^2}/{(2v_s^2)}-{v_t^2}/{(2v_s^2)}$ when tangential velocity
  is also available. The fact that the integral goes from $R_\mathrm{inner}$ to
  $R_\mathrm{max,t}$ indicates that the phase-space distribution at radius $r$
  has a contribution from tracers currently residing at larger radii, whose
  radial excursion includes  $r$.

\subsection{Likelihood and window function}
\label{sec:like}

The probability of each observed tracer object, labeled $i$, with
radius, $r_i$, radial velocity, $v_{ri}$, and tangential velocity,
$v_{ti}$, is
\begin{equation}
P_i(r_i,v_{ri},v_{ti}|\rho_s,r_s,\beta,\alpha,\gamma,r_0).
\end{equation}

Dynamical tracers, such as MW globular clusters, BHB stars and satellites, are
subject to selection effects. For example, sample completeness is often a
function of apparent magnitude (hence distance). If we assume that all
selection effects can be described by a window function, then the probability
of finding each tracer object, $i$, within the data window is given by the {\it
normalised} phase-space density \begin{equation}
  F_i=\frac{P_i}{\int_{\mathrm{window}} P \, \ud^3r \ud^3v}.  \label{eq:like}
\end{equation}
The integral in the denominator runs over the phase-space window. The
likelihood function then has the following form:

\begin{equation}
 L=\prod_i F_i.
\label{eq:like2}
\end{equation}

It can easily be shown that this likelihood function is equivalent to the
extended likelihood function marginalised over the amplitude parameter of the
phase-space density \citep[e.g.][]{1990NIMPA.297..496B}, which we are not
interested in. For our mock MW halo star catalogue, we deliberately exclude
stars in the innermost region of the halo. These stars have extremely high
phase-space density and so make a dominant contribution to the total
likelihood, strongly biasing the fit. We find that excluding all stars at
$r<7$~kpc removes this bias\footnote{A detailed discussion of the radial
dependence of results from our model is given in Sec.~\ref{sec:rbin}.}. The
window function in our analysis is then simply $P=0$ at $r<7$~kpc. In real
observations, the window function can be much more complicated.

We seek parameters that maximise the value of the likelihood function defined
in Eqn.~\ref{eq:like} and Eqn.~\ref{eq:like2}. In order to search the
high-dimensional parameter space efficiently, we use the software \textsc{IMINUIT},
which is a python interface of the \textsc{MINUIT} function minimiser
\citep{1975CoPhC..10..343J}.

There are six parameters in Eqn.~\ref{eq:radialonly2} or Eqn.~\ref{eq:RT2}. To
make best use of the likelihood method, we treat all six parameters as
free. In previous work using this approach the three parameters of the
spatial part of the tracer distribution are often fixed to their observed
values. We have carried out tests and found that, as expected, three parameter
models give results consistent with those using six parameters only when the
choice of tracer density profile is close to the true distribution. We
recommend that all six parameters should be left free if the observed sample
size is large enough to avoid introducing unnecessary bias. 

Another source of potential bias in the halo mass estimates of previous studies
arises from the use of universal mean mass--concentration relations for dark
matter halos. In CDM simulations, the relation between halo mass and
concentration has very large scatter \citep[e.g.][]{Neto_2007}. 
Taking halo A as an example, if we use the mass concentration relation 
from \cite{2008MNRAS.390L..64D}, the estimated concentration would be around 
5.7, which is almost three times smaller than the true value (see 
Table~\ref{tbl:parameters}). This would result
in an overestimate of halo mass by almost an order of magnitude, and the
corresponding scalelength, $r_s$, would be three times larger. The huge
scatter in the mass-concentration relation can cause catastrophic problems
unless we are fortunate enough that the host halo of the MW does in fact lie on
the mean mass-concentration relation.

\subsection{Weighting tagged particles}
\label{sec:weight}

As described in Sec.~\ref{sec:tagging}, our mock catalogues are created by
assigning stars from each single age stellar population to the 1\% most bound
dark matter particles in their host halo at the time of their formation. The
total stellar mass of each population will obviously vary from one population
to the next (according to our galaxy formation model), as will the number of
dark matter particles actually tagged (according to the number of particles in
the corresponding formation halo). The result is that stellar masses associated
with individual dark matter particles range over several orders of magnitude.
Particles tagged with larger stellar masses correspond to more stars, and thus
in principle should carry more weight in the likelihood fit.

To reflect this we could simply reweight each particle according to its
associated stellar mass, $M_{*,i}$. However, individual stars are \textit{not}
resolved: the phase-space coordinates of the underlying dark matter particles
comprise the maximum amount of dynamical information available from the tagging
technique. Therefore, we give each particle a weight
$({M_{*,i}}/{\Sigma_i{M_{*,i}}})N_\mathrm{tags}$. This conserves the total
particle number, $N_\mathrm{tags}$, but re-distributes this among particles in 
proportion to the fraction of the total stellar mass they represent. In this way 
we maintain a meaningful error estimated from the likelihood function. 

We also randomly divide stars into subsamples and apply our maximum
likelihood analysis to each of these to estimate the effects of Poisson noise.
To do so, we assign each weighted particle a new integer weight drawn from a
Poisson distribution with mean equal to the weight given by the expression
above. We repeat this procedure 10 times, so that we have 10 different
subsamples. The expectation values of the total weight for all tagged particles
in these subsamples are the same, so this approach can be regarded as analogous
to bootstrap resampling. We find this procedure yields consistent error 
estimates with that obtained from the Hessian matrix of the likelihood surface. 
From now on, we will only quote errors from the Hessian matrix.

We restrict our analysis to the 10\% oldest tagged particles in the main
halo. This is to reflect the fact that, in real observations, old halo stars
such as blue horizontal branch (BHB) and RR Lyrae stars are most often used as
dynamical tracers, because they are approximately standard candles.
We also exclude stars bound to surviving subhalos.

\subsection{Testing the method}
\label{sec:ideal}

\begin{figure}
\epsfig{figure=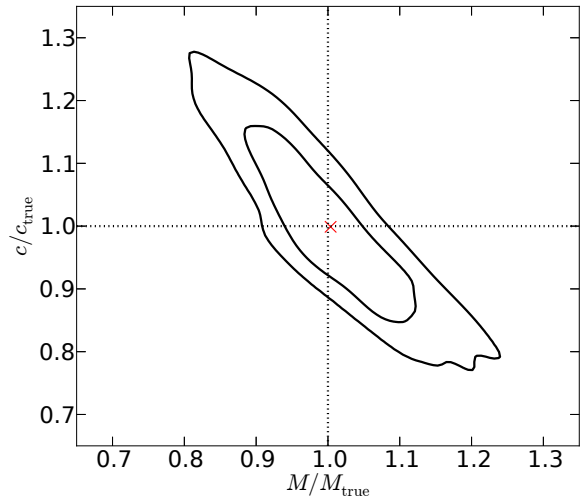,width=0.49\textwidth}%
\caption{The ratio between input and best-fit halo masses ($x$-axis) versus the 
ratio between input and best-fit halo concentrations ($y$-axis). Both radial 
and tangential velocities are used. The red cross is the mean ratio over 750 
different realisations, which is very close to 1 on both axes (horizontal and 
vertical dashed lines). Black solid contours mark the region in parameter plane 
enclosing 68.3\% (1$\sigma$) and and 95.5\% (2$\sigma$) of best fit parameters 
among the 750 realisations.}
\label{fig:ideal}
\end{figure}

Before fitting the model distribution function to our realistic mock stellar
halo catalogues, we test the method with ideal samples of particles that obey
Eqn.~\ref{eq:RT2}. We applied our maximum likelihood method to 750 sets of 1000
phase-space coordinates ($r$, $v_r$ and $v_t$) each drawn randomly from the same 
distribution function of the form given by Eqn.~\eqref{eq:RT2}. Fig.~\ref{fig:ideal}
shows a comparison between the input halo parameters and the recovered best-fit
halo parameters. The $x$ axis is the ratio between the best-fit and true-input
halo mass, and the $y$ axis the ratio between best-fit and true concentration.
The red cross indicates the mean ratios averaged over all the 750 realisations,
which is very close to unity (horizontal and vertical dashed lines). 

The best-fit halo mass and concentration varies among realisations
as a result of statistical fluctuations, as shown in
Fig.~\ref{fig:ideal}.  We note that the shape of these contours
indicate a correlation between the recovered halo mass and
concentration parameters. The correlation coefficient (i.e., normalised 
covariance) is -0.89, which implies a strong negative correlation
in the model parameter. We will discuss this correlation further 
in Sec.~\ref{sec:sourcesofbias}. The above exercise reassures us that 
our method works with ideal tracers, so we can move on to apply it to 
the more realistic mock halo stars in our simulations.

\section{Results}
\label{sec:6para}

In this section we investigate how well the true halo mass can be
recovered by fitting Eqn.~\ref{eq:radialonly2} to mock halo stars in
cases where: (a) only radial velocities are available
(Sec.~\ref{sec:radialonly}) and (b) both radial and tangential
velocities are available (Sec.~\ref{sec:RT}). In both cases we model
the underlying potential with infinite halo boundaries. We refer to
parameters estimated with the maximum likelihood technique as best-fit
(or measured) parameters, to be compared with the real (or true)
parameters taken directly from the simulation.

The total number of tagged particles we used in the five halos is shown in
Table~\ref{tbl:num}. These are of the order of $10^5$, one or two
orders of magnitude larger than the tracer samples used by 
\cite{2012MNRAS.424L..44D} or \cite{2012ApJ...761...98K}. This permits a robust
test of the method free from the effects of sampling fluctuations. Future
samples of observed tracers will grow with ongoing and upcoming surveys such
Gaia  \citep{2012AN....333..453P,2012Msngr.147...25G} and other deep spectroscopic 
surveys like MS-DESI \citep{2015AAS...22533605E} and 4MOST \citep{2012SPIE.8446E..0TD}.


\subsection{Radial velocity only}
\label{sec:radialonly}

\begin{figure}
\epsfig{figure=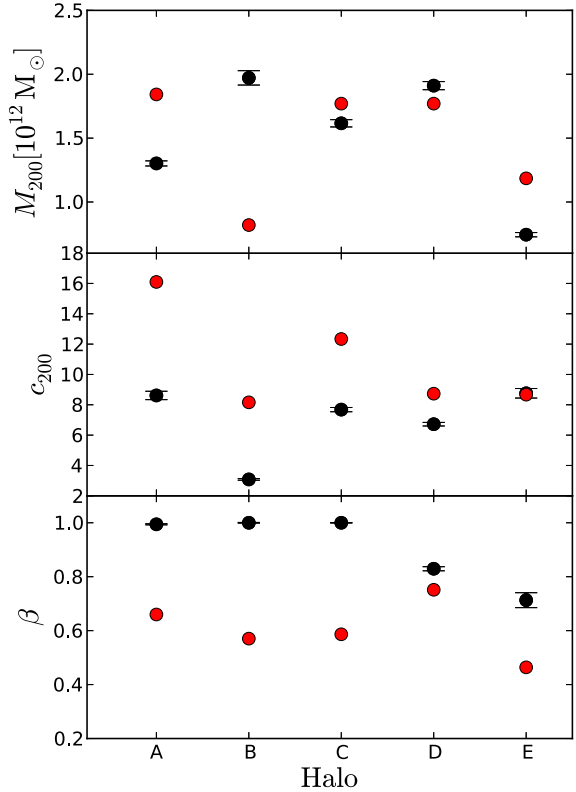,width=0.47\textwidth}
\caption{The best-fit values of $M_{200}$, $c_{200}$ and $\beta$ for the five 
halos (black dots with errors). Tangential velocities are not used. Error bars 
are 1$\sigma$ uncertainties obtained from the Hessian matrix and are almost 
invisible. The 1$\sigma$ errors are comparable to the scatter among the 10 
subsamples constructed in Sec.~\ref{sec:weight}. For direct comparison we show 
the true values of $M_{200}$, $c_{200}$ and $\beta$ as red dots. }
\label{fig:radialinfall}
\end{figure}

\begin{table*}
\caption{best-fit and true model parameters for each of our five
  halos. The highlighted rows list the true values of model
parameters and the subsequent two rows the corresponding best-fit
 values when using only radial velocities, $v_r$, and using both
radial and tangential velocities, $v_r+v_t$.
}
\begin{center}
\begin{tabular}{lrrrrr}
\hline
\hline
 & \multicolumn{1}{c}{A} & \multicolumn{1}{c}{B} & \multicolumn{1}{c}{C} & \multicolumn{1}{c}{D} & \multicolumn{1}{c}{E}\\ 
\hline
\rc true $M_{200} [\mathrm{10^{12} M_\odot}]$ &1.842&0.819&1.774&1.774&1.185 \\
 $v_r$ & 1.302 $\pm$ 0.02 & 1.972 $\pm$ 0.056 & 1.616 $\pm$ 0.029 & 1.911 $\pm$ 0.032 & 0.744 $\pm$ 0.017 \\
$v_r+v_t$ & 1.150 $\pm$ 0.003 & 0.867 $\pm$ 0.013 & 1.411 $\pm$ 0.013 & 1.410 $\pm$ 0.005 & 0.995 $\pm$ 0.014 \\
\rc true $c_{200}$ &16.098&8.161&12.337&8.732&8.667 \\
 $v_r$ & 8.616 $\pm$ 0.276 & 3.080 $\pm$ 0.053 & 7.682 $\pm$ 0.140 & 6.721 $\pm$ 0.118 & 8.758 $\pm$ 0.314 \\
$v_r+v_t$ & 15.269 $\pm$ 0.097 & 8.186 $\pm$ 0.107 & 15.878 $\pm$ 0.199 & 10.317 $\pm$ 0.039 & 10.297 $\pm$ 0.144 \\
\rc true $r_s[\mathrm{kpc}]$ &15.274&23.000&19.685&27.808&24.493 \\
 $v_r$ & 25.422 $\pm$ 1.006 & 81.660 $\pm$ 1.328 & 30.642 $\pm$ 0.494 & 37.035 $\pm$ 0.566 & 20.752 $\pm$ 0.636 \\
$v_r+v_t$ & 13.763 $\pm$ 0.088 & 23.360 $\pm$ 0.328 & 14.169 $\pm$ 0.183 & 21.805 $\pm$ 0.086 & 19.446 $\pm$ 0.287 \\
\rc true $\log_{10}\rho_s[\mathrm{M_\odot/kpc^3}]$ &7.193&6.591&7.025&6.646&6.642 \\
 $v_r$ & 6.664 $\pm$ 0.034 & 5.646 $\pm$ 0.016 & 6.544 $\pm$ 0.019 & 6.407 $\pm$ 0.018 & 6.681 $\pm$ 0.038 \\
$v_r+v_t$ & 7.278 $\pm$ 0.007 & 6.610 $\pm$ 0.014 & 7.321 $\pm$ 0.014 & 6.857 $\pm$ 0.004 & 6.852 $\pm$ 0.015 \\
\rc true $R_{200}[\mathrm{kpc}]$ &245.88&187.70&242.82&242.85&212.28 \\
 $v_r$ & 219.025 $\pm$ 11.155 & 251.525 $\pm$ 5.962 & 235.382 $\pm$ 5.786 & 248.901 $\pm$ 5.786 & 181.754 $\pm$ 8.579 \\
$v_r+v_t$ & 210.144 $\pm$ 1.797 & 191.234 $\pm$ 3.383 & 224.984 $\pm$ 3.785 & 224.962 $\pm$ 1.133 & 200.241 $\pm$ 3.770 \\
\rc true $\beta$ &0.660&0.570&0.587&0.752&0.464 \\
 $v_r$ & 0.994 $\pm$ 0.001 & 1.000 $\pm$ 0.001 & 1.000 $\pm$ 0.001 & 0.830 $\pm$ 0.008 & 0.713 $\pm$ 0.08 \\
$v_r+v_t$ & 0.458 $\pm$ 0.002 & 0.397 $\pm$ 0.002 & 0.407 $\pm$ 0.002 & 0.553 $\pm$ 0.001 & 0.254 $\pm$ 0.003 \\
\rc true $\alpha$  &2.926&2.912&3.055&2.007&2.223 \\
 $v_r$ & 2.965 $\pm$ 0.490 & 2.911 $\pm$ 0.007 & 3.008 $\pm$ 0.011 & 2.112 $\pm$ 0.009 & 2.454 $\pm$ 0.023 \\
$v_r+v_t$ & 2.774 $\pm$ 0.010 & 2.770 $\pm$ 0.008 & 2.962 $\pm$ 0.014 & 2.012 $\pm$ 0.005 & 2.413 $\pm$ 0.017 \\
\rc true $\gamma$ &6.468&7.485&6.383&6.048&5.256\\
 $v_r$ & 6.650 $\pm$ 0.037 & 8.362 $\pm$ 0.038 & 5.884 $\pm$ 0.034 & 6.031 $\pm$ 0.017 & 5.297 $\pm$ 0.023 \\
$v_r+v_t$ & 6.110 $\pm$ 0.025 & 8.140 $\pm$ 0.034 & 5.623 $\pm$ 0.030 & 5.820 $\pm$ 0.013 & 5.305 $\pm$ 0.020 \\
\rc true $r_0$[$\mathrm{kpc}$] &51.892&38.506&60.040&40.121&24.215\\
 $v_r$ & 53.376 $\pm$ 0.260 & 38.779 $\pm$ 0.138 & 57.643 $\pm$ 0.847 & 42.183 $\pm$ 0.204 & 26.645 $\pm$ 0.204 \\
$v_r+v_t$ & 42.590 $\pm$ 0.345 & 35.736 $\pm$ 0.111 & 51.375 $\pm$ 0.935 & 38.165 $\pm$ 0.100 & 26.645 $\pm$ 0.294 \\
\hline
\hline
\label{tbl:parameters}
\end{tabular}
\end{center}
\end{table*}

Fig.~\ref{fig:radialinfall} shows, as black points, the best-fit $M_{200}$,
$c_{200}$ and $\beta$ for our five halos in the case where only radial
velocities are known. These best-fit parameters are given in
Table~\ref{tbl:parameters} along with the true halo or tracer properties
(shaded in grey), which are plotted as red points in
Fig.~\ref{fig:radialinfall}.


\begin{table}
\caption{The total number of tagged particles in each of our five
  simulated halos.}
\begin{center}
\begin{tabular}{lrrrrr}\hline\hline
           & \multicolumn{1}{c}{A} & \multicolumn{1}{c}{B} & \multicolumn{1}{c}{C} & \multicolumn{1}{c}{D} & \multicolumn{1}{c}{E}\\ \hline
number & 181995 & 225030 & 184197 & 365280 & 120806 \\
\hline
\label{tbl:num}
\end{tabular}
\end{center}
\end{table}

Table~\ref{tbl:parameters} lists the true and best-fit values of the
host halo mass ($M_{200}$), halo concentration ($c_{200}$),
scalelength ($r_s$), scaledensity ($\rho_s$) and virial
radius ($R_{200}$).  Note only two of these parameters are independent. 
The best-fit $M_{200}$ values are
overestimates of the true values for halos B and D by 140\% and 7\%,
and underestimates for halos A, C and E by 30\%, 10\% and 35\%
respectively. Since the number of stars is very large (Table~\ref{tbl:num}), 
the statistical errors are all very small. The systematic biases in the estimates 
of halo mass and concentration are thus very significant. We find the scatter in 
parameters among the 10 bootstrap subsamples discussed in Sec.~\ref{sec:ideal} 
is comparable to the statistical error in the fit.


\begin{figure}
\epsfig{figure=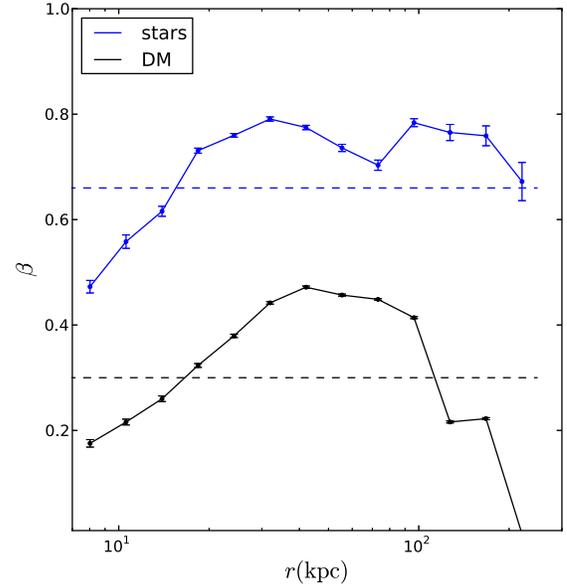,width=0.49\textwidth}
\caption{The velocity anisotropy parameter, $\beta$, as a function of
  radius for stars (blue solid curve) and a randomly selected
  subsample of dark matter particles (black solid curve) in halo
  A. Error bars are estimates obtained from bootstrap resampling. Blue
  and black dashed curves are the mean value of $\beta$ over the whole
  radial range for stars and dark matter particles. }
\label{fig:beta}
\end{figure}

The measured spatial parameters ($\alpha$, $\gamma$ and $r_0$) agree well with
the true values obtained from a double power-law fit to the stellar mass
density, shown as black dashed lines in Fig.~\ref{fig:densp}. The profiles
corresponding to the best-fit parameters are plotted as dashed green lines in
the figure. The agreement is especially good on scales smaller than
or comparable to the transition radius $r_0$. In the outskirts, differences
become noticeable for halos B and C. This is due to the fact that
there are fewer stars in these regions and the profiles have a significant
contribution from coherent streams.  As a result, the direct fitting of
radial profiles returns parameters that are slightly different from those
obtained from the likelihood technique because the latter also involves fitting
to the velocity distribution. In addition, the direct fit is dependent on our
choice of radial binning. 

The velocity anisotropy, $\beta$, is poorly estimated. The model
assumes $\beta$ to have a single value at all radii. However, the true
velocity anisotropy in the simulation does depend on radius: the blue
solid curve with errors in Fig.~\ref{fig:beta} is the velocity
anisotropy profile of stars in halo A as a function of distance from
the halo centre. We also show the mean value of $\beta$ (0.66 in
Table~\ref{tbl:parameters}) as the blue dashed line. The poor
measurement of $\beta$ is not simply due to radial averaging, because
we can see that the estimate of $\beta$ for halo A, 0.994, is
significantly greater than the real value over the whole radial range
probed.  The black curve shows the radial profile of $\beta$ for 
dark matter particles. There is an obvious offset between the velocity
anisotropy profiles of stars (tagged particles) and all dark matter,
which we will discuss in detail in Appendix~\ref{sec:betabias}.

\subsection{Radial plus tangential velocity}
\label{sec:RT}

\begin{figure}
\epsfig{figure=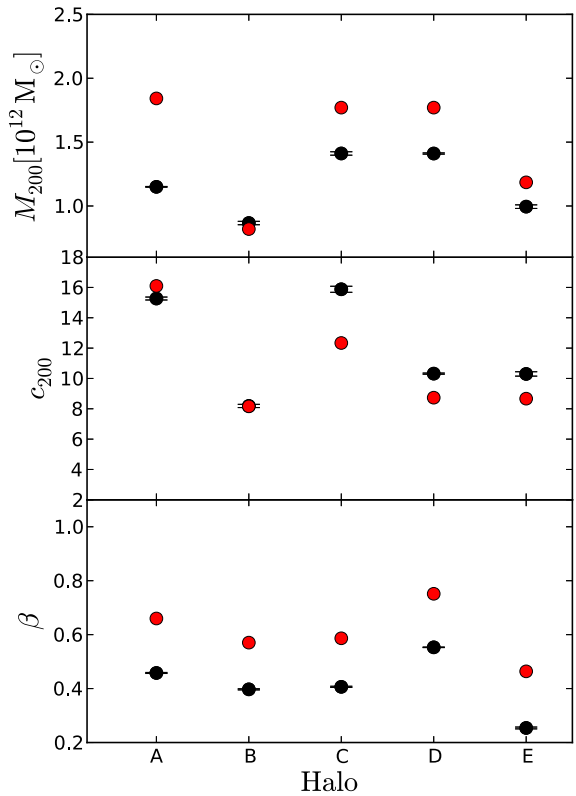,width=0.47\textwidth}
\caption{The measured $M_{200}$, $c_{200}$ and $\beta$. This is similar to 
Fig.~\ref{fig:radialinfall} but based on both radial and tangential velocities. 
Black solid dots with errors show our best-fit model parameters, while 
red dots show the true values of $M_{200}$, $c_{200}$ and $\beta$. }
\label{fig:RTinfall}
\end{figure}

Best-fit parameters when tangential velocities are also included are
shown as black points in Fig.~\ref{fig:RTinfall} and in
Table~\ref{tbl:parameters}. Compared with the results when only radial
velocities are used, we see a reduction in the overall bias of the
best-fit parameters with respect to their true values. However, there
are still significant discrepancies between best-fit and true
parameters, compared with the small errors. $M_{200}$ is
underestimated for halos A, C, D and E by 40\%, 20\%, 20\% and 15\%
respectively.  For halo B there is a 5\% overestimate.  Although the
measured host halo masses seem to be worse for halos A, C and D,
compared to the case where only radial velocities were used, the
agreement between measured and true halo concentrations in the same
halos is significantly improved. The best-fit spatial parameters, on
the other hand, converge to stable values that agree well with true
values.

Compared with Fig.~\ref{fig:radialinfall}, the measurements in
Fig.~\ref{fig:RTinfall} are much more clearly correlated with the true halo
properties. In particular, the shape of the best-fit and true $\beta$ curves
are in good agreement, although there is approximately a constant offset
between them. Tangential velocities are therefore essential for measuring
tracer velocity anisotropy, but even with this information there can still be a
systematic bias in the absolute value of $\beta$ recovered by distribution
function modelling. We return to this point in Appendix~\ref{sec:betabias}.

\subsection{Overall model performance}

\begin{figure}
\epsfig{figure=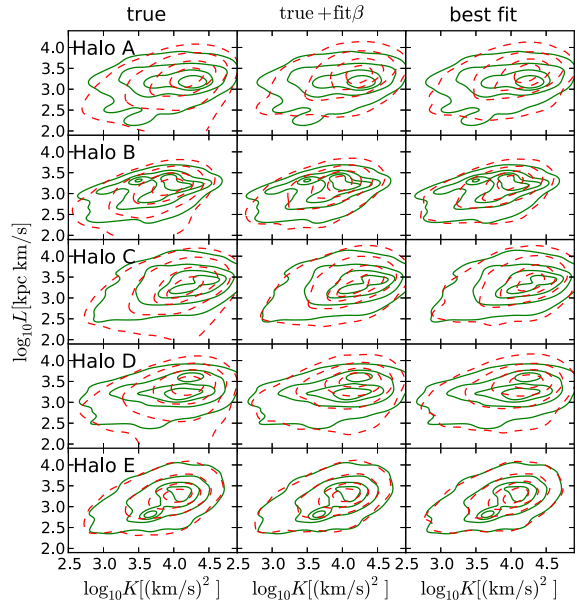,width=0.49\textwidth}%
\caption{A phase-space contour plot (kinetic energy, $K$, versus angular momentum, 
$L$) of mock halo stars (green solid curves) and model predictions (red dashed curves). 
The left column is based on true halo parameters, true $\beta$ and true spatial parameters 
of stars. The middle column is identical to the left column, except that $\beta$ has been 
fixed to its best-fit value in Table~\ref{tbl:parameters} (when both radial and tangential 
velocities are used). Halo and tracer parameters in the right column have all been fixed 
to be the best-fit parameters with both radial and tangential velocities. Contours for the 
five halos are presented in different rows, as indicated by texts in the left column of each 
row.}
\label{fig:modeloverall}
\end{figure}

We have shown that the degree of bias between true and best-fit values
resulting from our fitting procedure differs from halo to halo. In the
current subsection we aim to show how well the model works in
recovering the overall phase-space distributions of our mock halo
stars. Fig.~\ref{fig:modeloverall} shows phase-space contour plots for
mock halo stars (green solid curves) and compares them with the
predictions of our model (red dashed curves). We choose to make the
contour plot in terms of two observable quantities: kinetic energy, $K$, 
and angular momentum, $L$. Each row shows a different halo. In each
column, the choice of model parameters is different. In the leftmost
column, we use the true values of all parameters. In the case of 
$\beta$, we take its ``true'' value to be the velocity anisotropy 
averaged over the whole radial range. In the central column, we use the 
true values for all parameters \textit{except} $\beta$, which is set to 
be the best-fit value in Table~\ref{tbl:parameters} obtained using both 
radial and tangential velocities. In the rightmost column, we set all 
parameters to their best-fit values (again using both radial and tangential 
velocities). Since the green solid curves show data from the simulation, 
they are identical for all three columns of a given row. The contour 
levels correspond to the mass density of the 10, 30, 60 and~90\% densest 
cells.

The distribution functions defined by the true parameters (red dashed curves)
are a poor description of the mock stars in the left hand column, especially for
halos A, C and D where we see a significant over prediction of low angular
momentum particles. Halo E is the exceptional case in which we find good
agreement. The strongest disagreement in the other halos is, interestingly,
mainly due to the biased measurement of $\beta$. In the central column, where
we fix the value of $\beta$ to its best-fit value (obtained using both velocity
components) we see that the model predictions agree much better with the true
phase-space distribution, although some discrepancies remain. 

The fact that the model does not properly represent the distribution of 
mock stars when we use the true value of $\beta$ is indicative of possible 
deficiencies in the model functional form. We will show in Appendix A that 
the physical interpretation of $\beta$ in the power-law index, $-2\beta$, 
of our distribution function as the true velocity anisotropy is not appropriate. 
Moreover, the approximation of a constant $\beta$ over the whole radial range 
is problematic, as we know $\beta$ is radially dependent (see Fig.~\ref{fig:beta}). 
However, this is very likely subdominant because the true value of $\beta$ is 
above the best-fit value (0.458) over almost the entire radial range (blue 
solid curve in Fig.~\ref{fig:beta}).

For comparison, the right hand column of Fig.~\ref{fig:modeloverall} shows that
model predictions based on the best-fit parameters give an equally good match
to the simulation data. Judging by eye alone, it would be hard to tell whether
the middle column shows better or worse agreement than the right hand column.
However, judging according to the likelihood ratios, the best-fit halo
parameters are indeed a much better description of the data than the true halo
parameters ($\gg 3\sigma$). This is also reflected in the small formal errors
of the fit.  

\section{Sources of bias}
\label{sec:sourcesofbias}

Fig.~\ref{fig:modeloverall} indicates that the model is able to
recover the general phase-space distribution of the mock halo stars,
although there are some subtle factors which significantly bias our
best-fit parameter values relative to their true values. There are
several possible sources of this bias:

\begin{itemize}

 \item Correlations among parameters make the model more sensitive to
   perturbations and, in some cases, a poor fit to one parameter will propagate
   to affect the others. 
   
 \item Tracers may violate the assumption of dynamical equilibrium. 
 
 \item Both the underlying potential and the spatial distribution of tracers
   may not satisfy the spherical assumption.
 
 \item The velocity anisotropy ($\beta$) is not constant with radius as assumed 
   in the model, although this variation is probably subdominant compared 
   with the systematic bias in $\beta$.
 
 \item The true dark matter distribution may deviate from the NFW model.
 
 \item There are ambiguities in how to model the boundaries of halos.

\end{itemize}
 
We have investigated each of these possibilities and found that
ambiguity in the treatment of halo boundaries are relatively unimportant; 
hence we describe their effects in Appendix~B. We have investigated the density
profiles of the Aquarius halos and found that halo A is not well fit
by an NFW profile; instead its inner and outer density profiles are
better described by two different NFW profiles of different mass and
concentration. This might explain the systematic underestimation of
$M_\mathrm{200}$.  For the other halos, the NFW form is a good
approximation and thus deviations from it are not the dominant source
of bias. 

In Fig.~\ref{fig:RTinfall} we showed that the velocity anisotropy,
$\beta$, varies strongly with radius. The best-fit value of $\beta$
(which is assumed to be constant) turns out to show a significant
bias. We will discuss the origin of the bias for $\beta$ in Appendix~A. 
In the following, we will first discuss whether the bias and 
approximate treatment of $\beta$ affects the fit of the other 
parameters. In the following subsections, we focus on correlations 
among model parameters, the spherical assumption and the dynamical 
state of tracers.

\subsection{Correlations among model parameters}
\label{sec:cov}


\begin{table}
\caption{Correlation matrix of model parameters for halo E}
\begin{center}
\begin{tabular}{lrrrrrr}\hline\hline\\
\multicolumn{1}{c}{$v_r + v_t$}\\
\hline
  &\multicolumn{1}{c}{$\beta$}&\multicolumn{1}{c}{$M_{200}$}&\multicolumn{1}{c}{$c_{200}$}&\multicolumn{1}{c}{$\alpha$}&\multicolumn{1}{c}{$\gamma$}&\multicolumn{1}{c}{$r_0$}\\ 
\hline
$\beta$ & 1.0 & 0.025 & 0.033 & -0.00006 & -0.007 & 0.001 \\
$M_{200}$ & 0.025 & 1.0  & -0.887 & -0.207 & 0.342 & 0.040 \\
$c_{200}$ & 0.033 & -0.887 & 1.0 & 0.252 & -0.338 & -0.111 \\
$\alpha$ & -0.00006 & -0.207 & 0.252 & 1.0 & 0.382 & 0.768 \\
$\gamma$ & -0.007 & 0.342 & -0.338 & 0.068 & 1.0 & 0.724 \\
$r_0$ & 0.001 & -0.040 &  -0.111 & 0.768 & 0.724 & 1.0 \\
\hline\hline\\
\multicolumn{1}{c}{$v_r$ only}\\
\hline
&\multicolumn{1}{c}{$\beta$}&\multicolumn{1}{c}{$M_{200}$}&\multicolumn{1}{c}{$c_{200}$}&\multicolumn{1}{c}{$\alpha$}&\multicolumn{1}{c}{$\gamma$}&\multicolumn{1}{c}{$r_0$}\\ 
\hline
$\beta$ & 1.0 & -0.267 & -0.787 & 0.573 & 0.306 & 0.507 \\
$M_{200}$ & -0.267 & 1.0  & -0.321 & -0.244 & 0.187 & -0.073 \\
$c_{200}$ & -0.787 & -0.321 & 1.0 & -0.384 & -0.415 & -0.473 \\
$\alpha$ & 0.573 & -0.244 & -0.384 & 1.0 & 0.574 & 0.875 \\
$\gamma$ & 0.306 & 0.187 & -0.415 & 0.574 & 1.0 & 0.799 \\
$r_0$ & 0.507 & -0.073 & -0.473 & 0.875 & 0.799 & 1.0 \\
\hline\hline
\label{tbl:cov}
\end{tabular}
\end{center}
\end{table}

\begin{figure*}
\epsfig{figure=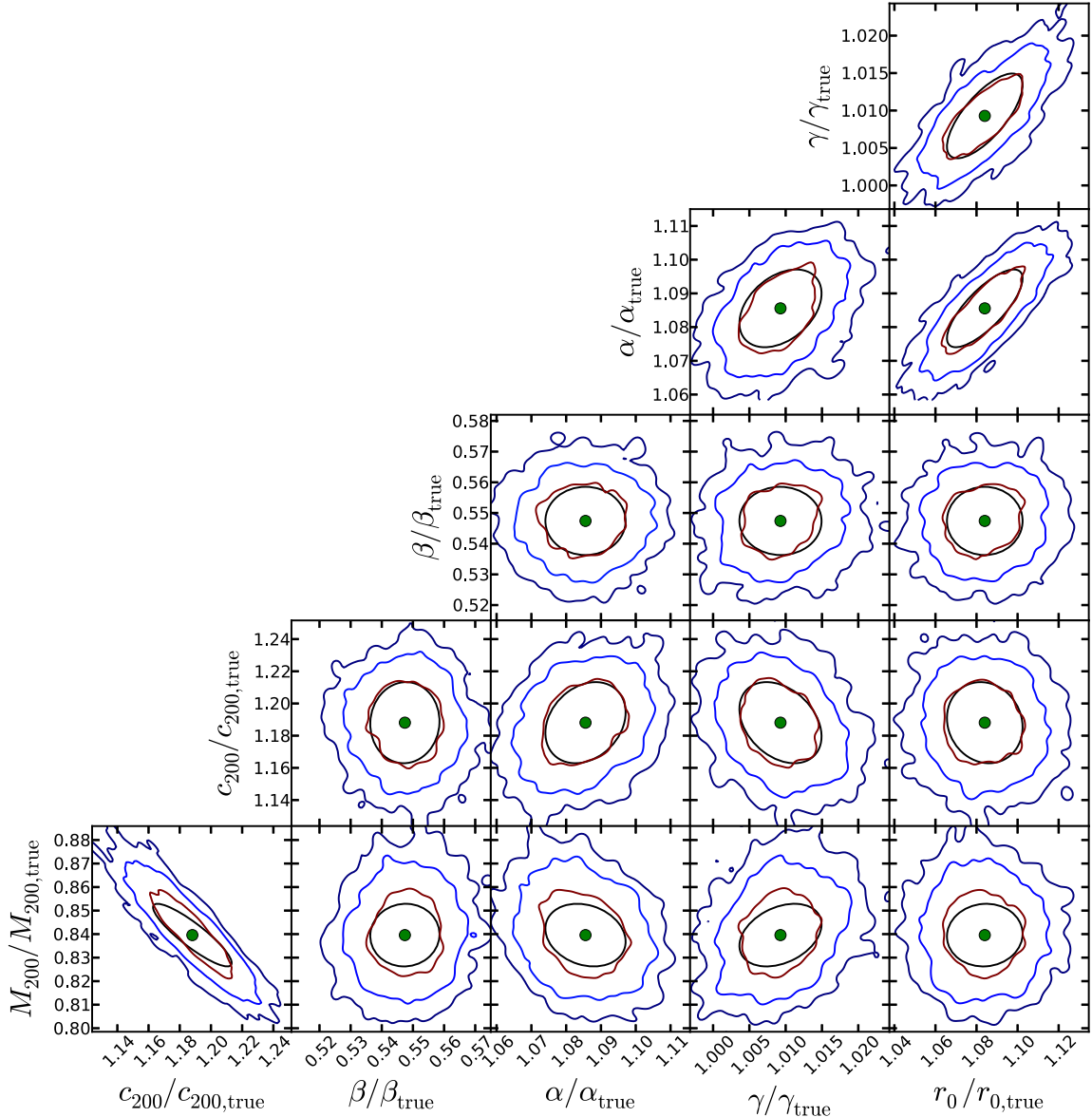,width=0.9\textwidth}%
\caption{The marginalised error contours for all possible combinations of every two out of 
the six model parameters (halo E).  Both radial and tangential velocities are used. Red, 
blue and purple contours correspond to 1, 2 and 3-$\sigma$ errors. Green dots show the 
best fit parameters scaled by their true values. The 1-$\sigma$ error ellipses estimated 
from the covariance matrix are over-plotted in black.}
\label{fig:contourTA}
\end{figure*}

\begin{figure*}
\epsfig{figure=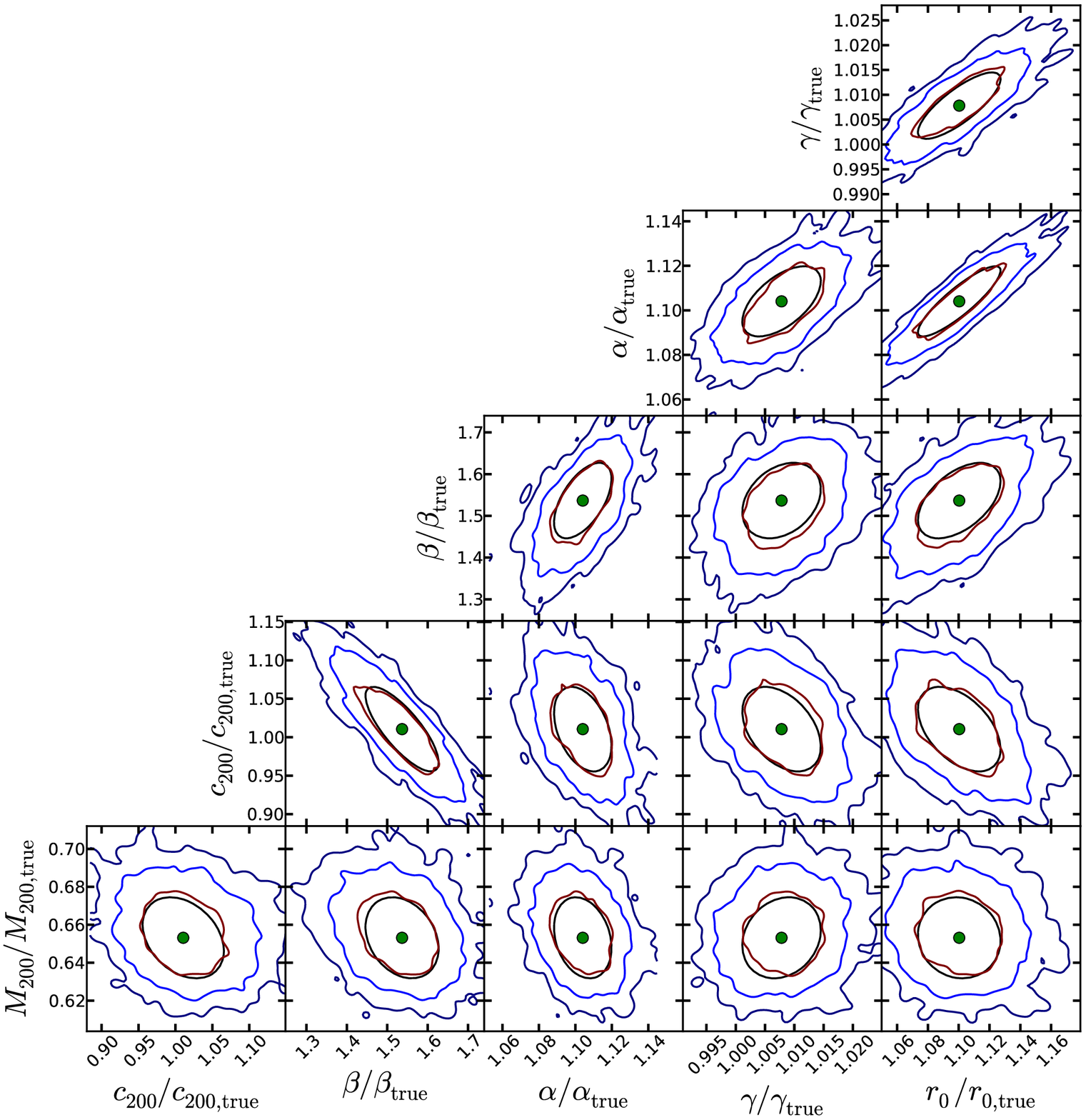,width=0.9\textwidth}%
\caption{As Fig.~\ref{fig:contourTA}, but only radial velocities are used.}
\label{fig:contourRA}
\end{figure*}

Fig.~\ref{fig:ideal} demonstrated a strong correlation between $M_{200}$ and
$c_{200}$. From a modelling perspective, this is dangerous: there are multiple
combinations of halo parameters that can give similarly good fits to both
the tracer density profile and velocity distribution. In this subsection we ask
what causes this correlation and whether there are similar correlations among
other parameters.  In particular, we have seen that the model gives strongly
biased estimates of the velocity anisotropy of stars, $\beta$. We want to check
whether this bias propagates to the other parameters.

Fig.~\ref{fig:contourTA} shows the marginalised 1, 2 and 3-$\sigma$ error contours for all 
possible combinations of two model parameters and is for halo E (tangential velocities 
are included as constraints). Fig.~\ref{fig:contourRA} is similar to Fig.~\ref{fig:contourTA}, 
but shows the corresponding error contours when only radial velocities are used. 
All parameters have been scaled by their true values (Table~\ref{tbl:parameters}). 
The error contours are obtained by scanning likelihood values over the full 6-dimensional 
parameter space. We also overplot as black ellipses the 1-$\sigma$ error from the 
covariance matrix recovered for halo E. The agreement between the black ellipses 
and blue error contours is very good in all the panels, indicating the error estimated 
from the Hessian matrix is robust. The corresponding values of the normalised 
covariance matrix are also provided in Table~\ref{tbl:cov}.

For the other halos, the error contours look qualitatively similar, except for 
halo A in which the correlation between halo mass and concentration is weaker. This 
is because the bias in the recovered halo properties of halo A is mostly due to its 
deviation from an NFW profile. Otherwise, we found the agreement between the error 
ellipse from the covariance matrix and the scanned error contours are worse for the 
radial velocity only case of halo A, B and C. This is mainly because the best fit value 
of $\beta$ touches the upper boundary $\beta=1$ and thus the quadratic approximation is 
no longer good enough to estimate the error from Hessian matrix, but even in such cases 
the errors estimated from Hessian matrix is still acceptable with at most a factor 
of two deviation from the more strictly obtained error contours.

From Fig.~\ref{fig:contourTA}, Fig.~\ref{fig:contourRA} and Table~\ref{tbl:cov}, 
we can see the correlation between $M_{200}$ and $c_{200}$ is very strong when 
including tangential velocities (covariance close to -1). The correlation is not 
as strong if only radial velocities are used.  To help understand this correlation,
we have explored the velocity distribution of tracers predicted by the model using
different combinations of $M_{200}$ and $c_{200}$. We verify that, if $M_{200}$ is
increased, the predicted velocity distribution of stars extends to larger velocities, 
with a corresponding reduction in the probability of smaller velocities. A decrease 
in $c_{200}$ can roughly compensate for this change in the velocity distribution.

It is worth emphasising that although the error contours for $M_{200}$ and 
$c_{200}$ are highly elongated (corresponding to the correlation between the two), 
they are still closed, indicating the constraining power is not insignificant. 
Because these contours represent the statistical error, they can be reduced by 
increasing the sample size. With our current sample of ~$10^5$ particles, 
the statistical errors on $M_{200}$ and $c_{200}$ are controlled to the 1\% level, 
which is negligible compared to the systematic bias in the parameter values. 
In other words, the true $M_{200}$ and $c_{200}$ values lie well outside the 3$\sigma$ 
confidence contour, so that statistical fluctuations do not explain the deviation 
between the fit and the true parameters even after considering the correlation.

It is interesting to see that the best fit $M_{200}$ and $c_{200}$ in Fig. 6
tend to be biased in opposite directions, except for halo A. Such biases are
mainly systematic, because the statistical errors are much smaller. This
indicates a negative correlation between the systematic biases, similar to the
statistical correlation we have seen above. Note that in principle the
correlation in the systematic biases could happen along any direction,
independent of the statistical correlation, and it is unclear why the two act
in the same direction here. A clean and thorough exploration of this involves
segregating various model assumptions and adopting a large sample of halos;
\cite{han2015a,han2015b} present part of this work, which will also be
discussed in a forthcoming paper by Wang et al. (in preparation). At this
stage, we provide some further discussions on this point in our conclusion.

\begin{figure}
\epsfig{figure=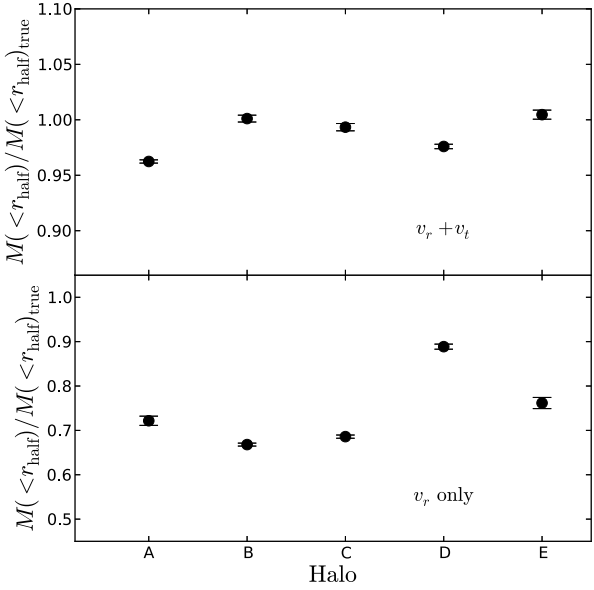,width=0.5\textwidth}%
\caption{The ratio between the best fit and true mass enclosed within half-mass 
radius. Errors are obtained through error propagation from the covariance matrix 
of $\rho_s$ and $r_s$, with the correlated error between $\rho_s$ and $r_s$ 
included.}
\label{fig:masshalf}
\end{figure}

Similar correlation
between the velocity anisotropy parameter, $\beta$, and halo properties have 
been discussed in some previous studies of the Milky Way and dwarf galaxies 
\citep[e.g.][]{2009ApJ...704.1274W,2010MNRAS.406.1220W, 2013JCAP...07..016N}, 
though their models are different. In particular, \cite{2009ApJ...704.1274W} 
and \cite{2010MNRAS.406.1220W} have reported the mass within the 
half light radius of dwarf galaxies is relatively insensitive to the value 
of $\beta$. Here we have also tested whether our model can better constrain 
the mass within the half-mass radius\footnote{The half-mass radius is 
defined to be the radius inside which the enclosed stellar mass is half of 
that of the whole tracer population.} of our stellar tracers, and the results 
are shown in Fig.~\ref{fig:masshalf}. 

The two panels of Fig.~\ref{fig:masshalf} show results based on 
both radial and tangential velocities (top) and radial velocities only 
(bottom). Black dots with errors are the ratio between the best fit mass 
and the true mass within the half mass radius. Encouragingly, we see a very 
good agreement between the best fit and true mass if tangential velocities 
are used. The levels of biases are about 3.8\%, 0.7\% and 2.4\% underestimates 
for halos A, C and D. For halos B and E the mass is overestimated by 0.1\% 
and 0.5\% respectively. On the other hand, if only radial velocities are 
used the bias is much more significant. We underestimate the mass by about 
27.8\%, 33.2\%, 31.4\%, 11.1\% and 23.8\% for the five halos. Compared with 
Fig.~\ref{fig:radialinfall}, the level of bias becomes significantly smaller 
for halo B, and is slightly improved for halos A, C and D. Our results suggest 
if tangential velocities are available, the mass within a fixed radius close 
to the half-mass radius of tracers are not sensitive to the parameter correlation 
and can be constrained much better than the total halo mass.

\begin{figure}
\epsfig{figure=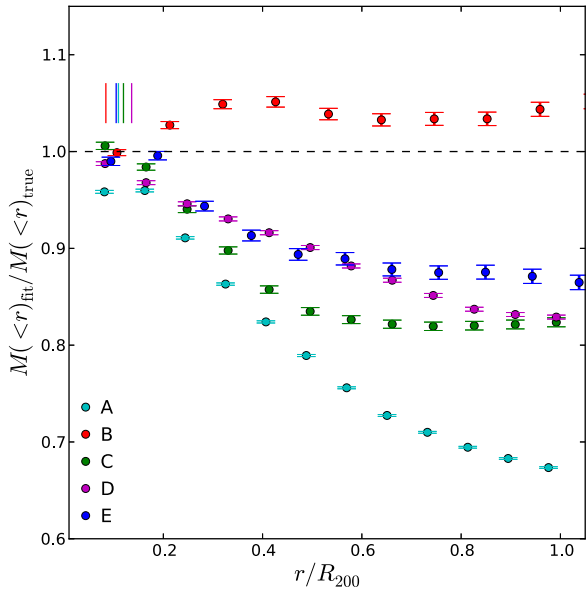,width=0.5\textwidth}%
\caption{The best-fit total mass within a fixed radius compared to the
  true mass within that radius. Both radial and tangential velocities 
  are used. Errors are obtained through error propagation from the 
  covariance matrix of $\rho_s$ and $r_s$, with the correlated error 
  between $\rho_s$ and $r_s$ included. The black dashed line marks 
  equality between the measured and true mass. }
\label{fig:massinside}
\end{figure}

In Fig.~\ref{fig:massinside}, we examine the halo mass profiles (cumulative mass 
within a certain radius as a function of the radius) with the best-fit parameters 
(when both radial and tangential velocities are included), normalised 
by profiles with the true parameters. Except for halo B, which gives an acceptable 
result at all radii, the measurements are very close to the true value for $r\leq 0.2R_{200}$ 
with a less than 5\% bias, though the bias is still significant given the small statistical 
errors. The measurements become significantly more biased at larger radii. The vertical  
lines mark the locations of half-mass radii of stellar tracers in the five halos, which 
are close to $0.1R_{200}$. In practice this means the mass enclosed within 40 to 60 
kpc can be more robustly determined as it suffers less from the correlation, and 
this is usually the radial range for which relatively large numbers of tracers can be 
observed. Thus the mass measurements reported within 40 to 60~kpc in the literature are 
expected to be more robust. Note, however, the result here is obtained from stars over 
the whole radial range. We will further show in Sec.~\ref{sec:rbin} that the mass within 
about $0.2R_{200}$ can also be determined robustly if stellar tracers are restricted to 
be within 60~kpc. 

In contrast to the strong correlation between $M_{200}$ and $c_{200}$, the
correlation between $\beta$ and all the other parameters is very weak if tangential 
velocities are included. This is fortunate, as it suggests the systematically biased 
estimate of $\beta$ will not introduce further bias to the other parameters if we have 
the proper motion information. On the other hand and for the radial velocity 
only case, the correlation between $\beta$ and the other parameters are actually quite 
strong, especially in terms of the correlation with halo concentration. This suggests 
if one only has radial velocity information, it is hard to get a robust estimate of 
$\beta$ (Fig.~\ref{fig:radialinfall}) and the bias in $\beta$ 
may affect the fitting of the other model parameters. Compared with the systematic  
bias in $\beta$, the radial dependence of $\beta$ is actually subdominant.

Correlations between halo parameters ($M_{200}$ or $c_{200}$) and tracer spatial
parameters ($\alpha$, $\gamma$ and $r_0$) are at the level of a few tens of
percent. For the case when tangential velocities are included, an increase in 
the tracer density outer slope would cause an increase
in the recovered halo mass and a corresponding decrease in halo concentration;
conversely, an increase in the inner slope would cause a decrease in the halo
mass and an increase in the concentration. As a result, uncertainties in the fit 
to the tracer density profile may further bias the best fit halo parameters.  
For example, the best-fit (green dashed) curve in the
halo C panel of Fig.~\ref{fig:densp} agrees well with the true profile (red
points) inside 170~kpc but is somewhat shallower at larger radii. If we fix the
three spatial parameters in our fit to halo C to those given by a conventional
reduced-$\chi^2$ best-fit to the tracer density (dashed black curve in
Fig.~\ref{fig:densp}) the best-fit halo mass is boosted by about 10\%. If the
tracer density profiles deviate from the double power-law form, these
correlations between halo parameters and spatial parameters would introduce
further bias to the best-fit halo mass.  

Lastly, we note that correlations between the three spatial parameters are strong 
as well. This quantifies our earlier finding that, in the case of halo A, adding
tangential velocities as constraints in the fit makes the outer slope of the
tracer density profile shallower and the break radius smaller, but results in
very little perceptible change in the overall profile shape. Hence, good fits
to the tracer density profiles may not be unique.  An increase in $r_0$ can be
roughly compensated by a corresponding increase of both $\alpha$ and $\gamma$.


\subsection{Model uncertainties in the spherical assumption}
\label{sec:triaxis}

\begin{figure}
\epsfig{figure=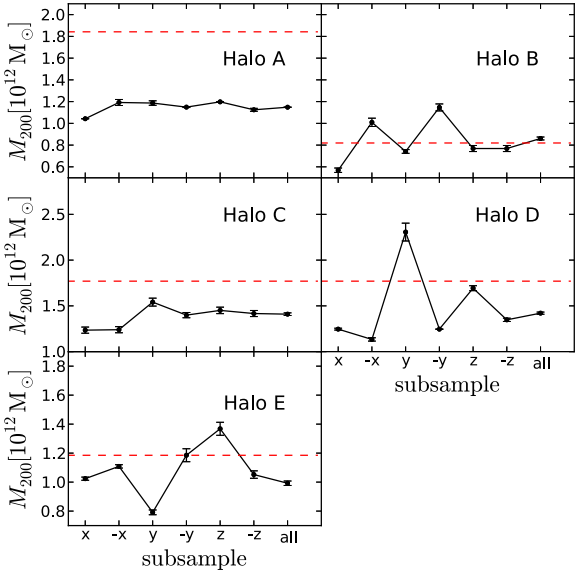,width=0.49\textwidth}%
\caption{The host halo masses obtained by fitting to samples of stars drawn from 
different sky directions and using both radial and tangential velocities. Halos 
have been rotated based on their dark matter distribution. The $x$-axis and $z$-axis 
in the rotated system are chosen to be the major and minor principles of the dark 
matter halo. Measurements are repeated for six survey cones centred at $\pm x$, 
$\pm y$ and $\pm z$  direction, with opening angle $\pm \frac{\pi}{4}$. Results 
based on all stars are plotted as the right most point. The red dashed lines 
show the true values of $M_{200}$.}
\label{fig:rotation}
\end{figure}	

In our analysis and the majority of existing studies of using dynamical tracers
to constrain the MW host halo mass, both the tracer population and the
underlying potential are modeled assuming spherical symmetry. 
However, we know dark matter halos in N-body simulations are triaxial
\citep[e.g.][]{2002ApJ...574..538J}, and the spatial distribution of tracers is
unlikely to be perfectly spherical either. It is thus necessary to investigate
how the triaxial nature of the underlying dark matter potential and tracer
populations affect our results.

To do this, we first rotate the five halos to a new Cartesian coordinate
system defined by their principle axes. In this rotated system, the $z$-axis is
aligned with the minor axis of the halo and the $x$-axis with major axis of the
halo. We then repeat our analysis using six subsamples of tracers drawn from
mock `survey' cones pointing along each of the three axes from the origin at
the centre of the halo, in the positive and negative directions. The opening
angle of each cone is $\pm \frac{\pi}{4}$.

Fig.~\ref{fig:rotation} shows the recovered host halo masses for each of the six
cones. Tangential velocities are included in the fit. For a direct comparison,
we have also plotted results based on all tracers, as the right most point
(from Table~\ref{tbl:parameters}). There are significant variations in the
results obtained from surveys along different directions, ranging from only a
few percent (halo A) to as much as a factor of two (halo D and E). Halos D and
E have the most obvious variations. We have explicitly checked that the
significant overestimate along the positive $y$-axis of halo D is due to the
existence of four relatively massive subhalos
($M_\mathrm{subhalo}/M_\mathrm{200,host}>1\%$), while the large variation in
halo E is due to one prominent stream (see the yellow dots in the bottom panel
of Fig.~\ref{fig:scatter} or Fig. 6 in \cite{2010MNRAS.406..744C}), which
ranges from $r\sim 80$~kpc ($\sim 0.3 R_\mathrm{200}$) all the way to the virial
radius. 

These variations are, however, almost random and uncorrelated with the choice
of any particular principle axis, and they change from halo to halo. Halo A has
the smallest variation, with all results well below the true host halo mass
(red dashed line). Although stronger variations are seen for halo C, all
results are again well below the true host halo mass. Thus despite the fact
that the variations between directions can be as large as a factor of two, this
does not seem to be the dominant cause of the systematic differences between
the best-fit and true halo mass in our analysis.

\subsection{Unrelaxed dynamical structures}
\label{sec:stream}

\begin{figure*}
\epsfig{figure=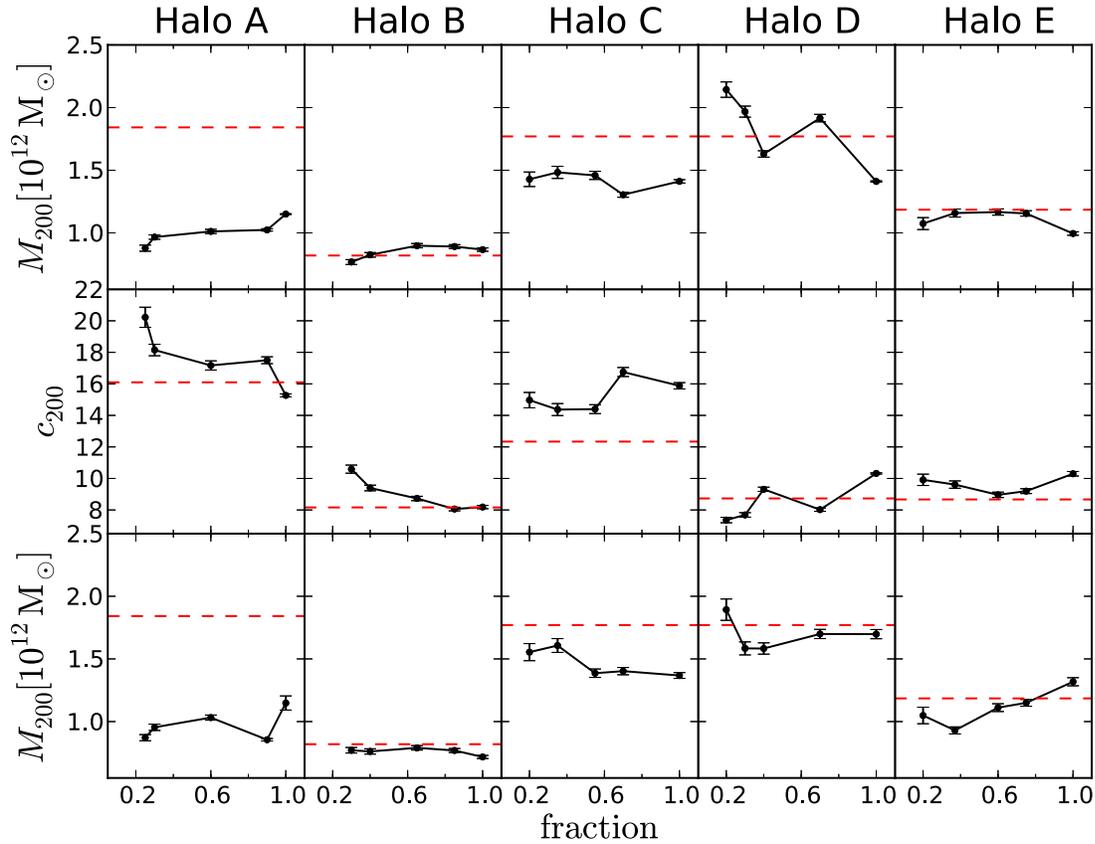,width=0.9\textwidth}%
\caption{The host halo masses ($M_{200}$) and concentrations
  ($c_{200}$) measured using a certain fraction of stars that have the
  earliest infall time and with both radial and tangential velocities.
  The five columns are for five Aquarius halos,
  as labeled at the top. The red dashed lines show the true values of
  $M_{200}$ and $c_{200}$. The top and middle rows are based on 
  the 10\% oldest tagged particles. The bottom row is analogous to the
  top one, except that stars whose parent satellites
  have not been entirely disrupted yet are excluded.}
\label{fig:nosat}
\end{figure*}

\begin{figure*}
\epsfig{figure=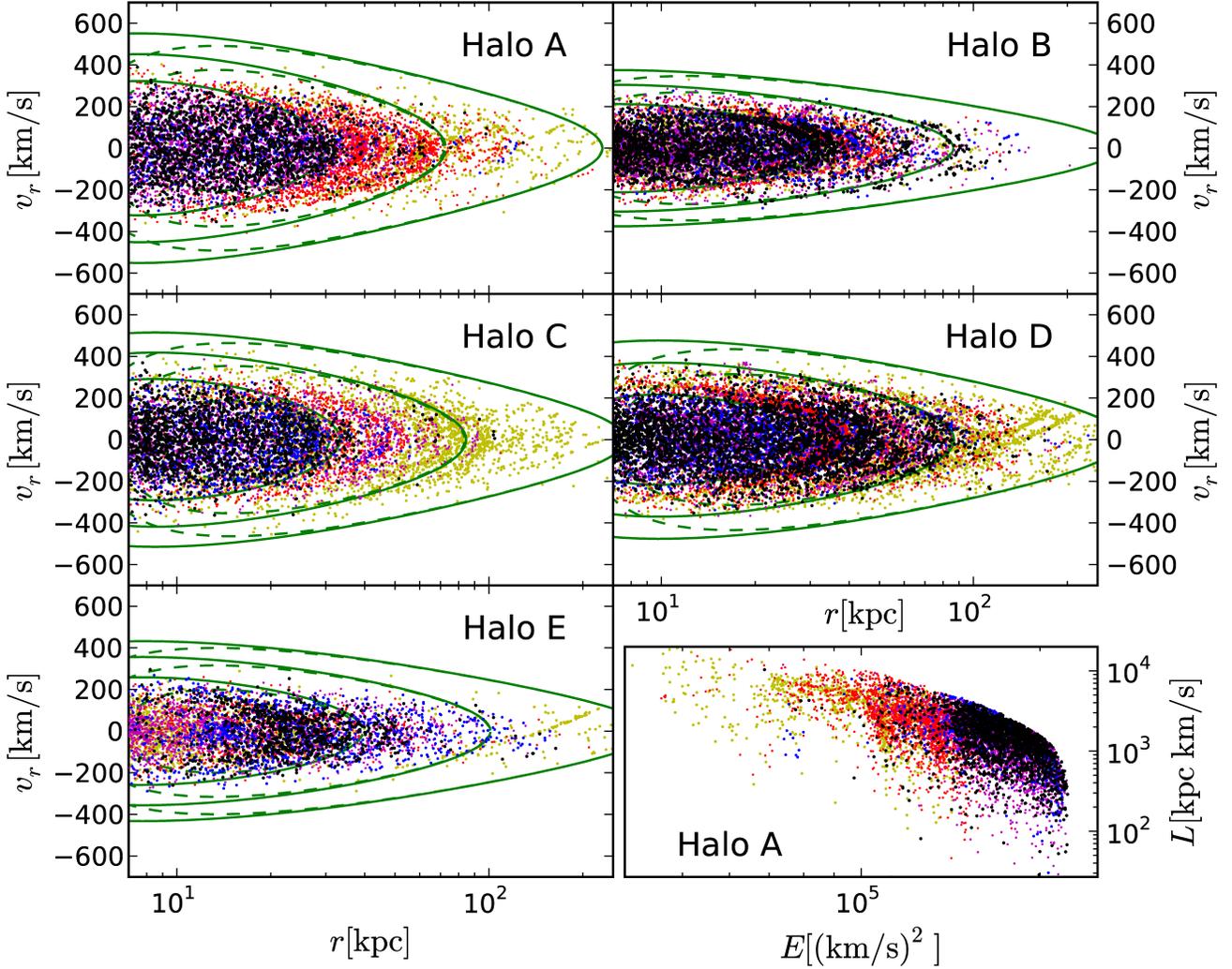,width=1.\textwidth}%
\caption{Phase-space scatter plot (radial velocities versus radial positions) 
for the 10\% oldest tagged particles in the five Aquarius halos. Data points are 
colour-coded according to their infall time. In sequence, points with yellow, red, 
magenta, blue and black colours are stars that have earlier and
earlier infall times. The bottom right panel is a scatter plot of
energy and angular momentum for halo~A. Only one out of every 20 points are 
plotted in order to avoid saturation. Green curves are contours of constant angular 
momentum and binding energy (see text for details).}
\label{fig:scatter}
\end{figure*}

\begin{figure}
\epsfig{figure=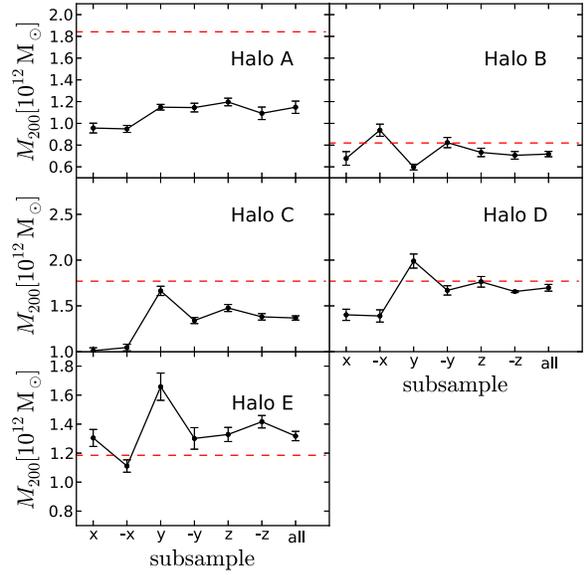,width=0.49\textwidth}%
\caption{Similar to Fig.~\ref{fig:rotation}, but only using stars whose parent 
satellites have been entirely disrupted.}
\label{fig:rotationnosat}
\end{figure}	

The model distribution function used in our analysis assumes that the tracer
population is in dynamical equilibrium and hence the phase-space density
of tracers is conserved. Our mock halo stars are all accreted from satellite
galaxies, with a range of accretion times. Some prominent phase-space
structures, such as stellar streams, may therefore violate the assumption of
dynamical equilibrium. In this section we ask how the presence of unrelaxed
dynamical structures affects our results.

We expect the dynamical state of stars in our mock catalogue to depend on the
infall redshift of their parent satellite, at least approximately (satellites
on different orbits will have different rates of stellar stripping). We might
expect to obtain an improved mass estimate if we use only stars from satellites
that fell in earlier, because they have had more time to relax in the host
potential. To test this, we rank halo stars according to their infall
time\footnote{Defined as the simulation output redshift at which the parent
satellite of each star reaches its maximum stellar mass, which is generally
within one or two outputs of infall as defined by SUBFIND.}.  We measure the
host halo mass and concentration with samples defined by different cuts in
infall time using both radial and tangential velocities, corresponding 
to roughly the same fraction of stellar mass in 
each halo. The top and middle rows of Fig.~\ref{fig:nosat} present these
parameters as a function of the fraction of stars selected by each cut, for the
five different halos.  A small fraction corresponds to an earlier mean infall
time, but also (obviously) to a smaller sample size.  A fraction of $1$ means
all the mock stars have been included, hence the corresponding parameters are
those listed in Table~\ref{tbl:parameters}. 

We see fluctuations in the measured halo properties with infall time, but no
obvious trends. Using samples of stars with earlier mean infall times does not
seem to reduce the bias between best-fit and true parameters.  This may be
because the dynamical state of tracers depend on many other factors, such as
the orbit of their parent satellites\footnote{We have carried out an analogous
exercise in which we rank stars by the time at which they are stripped from
their parent satellite. We found that this stripping time correlates with the
infall time of the parent satellite, and the conclusions regarding the
recovered halo parameters are similar.}. Samples for which the measured halo
masses increase produce a corresponding decrease in the measured
concentrations, again reflecting the strong correlation between $M_{200}$ and
$c_{200}$.  

To gain more intuition regarding the dynamical state of halo tracers,
Fig.~\ref{fig:scatter} shows phase-space scatter diagrams for mock halo stars
(radius, $r$, versus radial velocity, $v_r$).  Points are colour coded according
to the infall time of their parent satellite, with black points corresponding
to satellites falling in earliest and blue, magenta, red and yellow points to
successively later infall times. Stars with earlier infall times are clearly
more centrally concentrated. For points in Fig.~\ref{fig:RTinfall} with 
decreasing fraction of stars that fell in earliest, the corresponding particles 
in Fig.~\ref{fig:scatter} can be found by excluding yellow, red, magenta 
and blue points by sequence and looking at the remaining points.

Green curves in Fig.~\ref{fig:scatter} are contours of constant angular momentum
and binding energy. There are six contours in total, corresponding to three
discrete values of binding energy and two discrete values of angular momentum:
dashed lines have a higher angular momentum than solid lines. Smaller maximum
radius indicates higher binding energy. It is thus straightforward to see that
particles with higher binding energy have smaller velocities and are more
likely to be found in the inner regions of the halo. Comparing the solid and
dashed contours, we see that increasing angular momentum at fixed binding
energy causes significant differences in the inner regions of the halo, while
at larger radii the two sets of contours almost overlap.

We can see that points with the same colour trace these contours with some scatter,
implying that stars whose parent satellites fall in at a particular epoch share
similar orbits. This can be seen more clearly in the bottom right
panel, which shows a scatter plot of binding energy versus angular momentum for
stars in halo A.  Points with the same colour occupy regions covering a narrow
range of binding energy. The correlation between infall time and binding energy
of subhalos has been studied by \cite{2012MNRAS.425..231R}. {\it Here we have shown
that stars from stripped subhalos show a correlation between infall redshift
and binding energy as well.}

Although mock stars trace the green contours overall, we can still see some
prominent structures. For example, there are two yellow spurs in the outskirts
of halo D and one yellow spur in halo E. These correspond to particles that
have only just been stripped from their parent satellites. These stars are far
from equilibrium: their exclusion causes the rapid change in $M_{200}$ and
$c_{200}$ in Fig.~\ref{fig:nosat} between fractions of 1 and $\sim0.7$ in
halos D and E. 

To confirm that these unrelaxed phase-space structures can affect our
results, we have repeated the above exercise excluding all stars whose
parent satellites have not been entirely disrupted. Corresponding
results are shown in the bottom row of Fig.~\ref{fig:nosat}, again
ranking stars by their infall time. Measured halo masses are clearly
affected by excluding stars whose parent satellites still survive. For
halos A and C, we see some small fluctuations in the measured halo
mass, but the systematic underestimate of the true halo mass still
remains. The most dramatic changes occur for halos B, D and E. First,
the point corresponding to a fraction of 1 for halos D show a
significant increase in the recovered mass towards the true values,
reinforcing our conclusion that unrelaxed structures are causing
significant underestimates of $M_{200}$ in these halos. In fact,
\textit{most} of the yellow dots in halo D panel of
Fig.~\ref{fig:scatter} are stars that have been stripped from
satellites that still survive. After excluding these, the two
highest-fraction points in the halo D panel are based on almost the
same sample of stars. We also notice that fluctuations around the true
value for the different fractions are reduced in the bottom row (for
example, the two lowest fraction points in halo D). 

The effects of excluding halo stars from surviving satellites are more 
ambiguous in halos B and E. The recovered mass of halo B decreases slightly, 
while for halo E the rightmost point, corresponding to all stars, increases, 
but the two left most points decrease in amplitude, causing 
a stronger deviation away from the true values.

We further investigate how the measurements pointing in different directions 
change with more relaxed stars. Fig.~\ref{fig:rotationnosat} repeats 
Fig.~\ref{fig:rotation}, using only those stars from satellites that have been 
entirely disrupted. The measured $M_{200}$ of halo A shows some small variations 
compared with Fig.~\ref{fig:rotation}, but the variation is too small to be 
significant. The recovered mass of halos B and C improves in some directions, 
whereas in some other directions it worsens. The most encouraging improvements are 
for halos D and E. The two measurements along $\pm x$ directions of halo D remain 
almost unchanged, while the measurements in all the other directions are significantly 
improved. For halo E the recovered mass increases significantly in all directions.

Our conclusion is thus for halo D (and perhaps E) the underestimates of their host halo 
masses when all particles are used are mainly due to unrelaxed dynamical structures; for 
the other halos, the effects of unrelaxed dynamical structures are not obvious.  
Stars with surviving parent satellites in halos A, B and C could be more dynamically 
relaxed and thus excluding stars that are expected to be unrelaxed does not make a 
significant change. Furthermore, we note in a recent work, \cite{2014ApJ...795...94B} 
has developed a new method of determining halo potential using tidal streams. They 
found individual streams can both under- and overestimate the mass, but the whole stream 
population is essentially unbiased. Though their method is different from ours, 
it is possible that a single dynamically hot stream can potentially bias the result, 
the combination of several streams can help to bring an overall unbiased measurement. 
A more detailed study quantifying the dynamical state of tracers has been carried 
out in \cite{han2015a,han2015b}.

\section{Model uncertainties in the radial average and implications for real surveys}
\label{sec:rbin}

\begin{figure}
\epsfig{figure=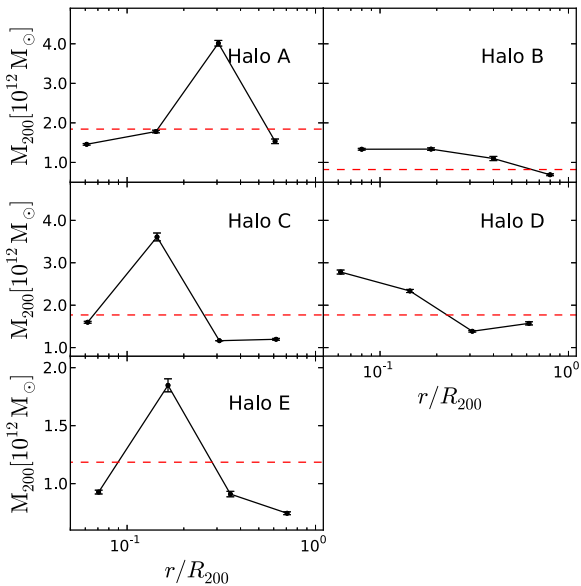,width=0.49\textwidth}
\caption{Best-fit host halo masses using samples of stars in four
  radial bins, (7-20)~kpc, (20-50)~kpc, (50-100)~kpc and $>$100~kpc. 
  Both radial and tangential velocities are used.} 
\label{fig:rbin}
\end{figure}

\begin{figure}
\epsfig{figure=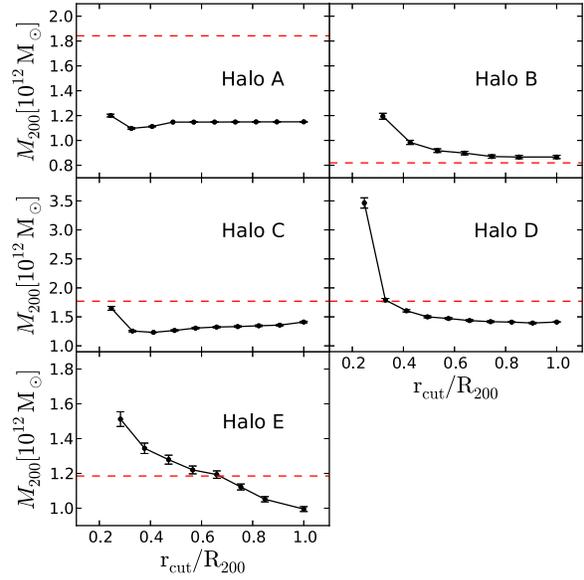,width=0.49\textwidth}%
\caption{best-fit host halo masses as a function of the outer radius limit of the tracer 
population ($r<r_\mathrm{cut}$). Both radial and tangential velocities are used.}
\label{fig:rcut}
\end{figure}

We have seen in the previous section that our maximum likelihood technique
recovers different halo mass from sets of tracers with different infall
redshifts, or more fundamentally,  different binding energies. The sense and
magnitude of these differences show no obvious correlations with either
quantity, however. Stars falling in earlier typically have high binding energy
and are mostly concentrated in the central regions of the halo; since binding
energy correlates with radius, we may also expect fluctuations in the recovered
halo parameters when using samples of stars drawn from a particular radial
range. In this section we investigate this radial dependence. This helps to
understand the behaviour of the full model, which averages over all radii.
Variations with radius are relevant to observational applications as well,
because in practice tracers are often selected from relatively narrow radial
ranges, and these ranges may be different for different tracers.

We assign stars to four bins of galactocentric radius: (7-20)~kpc,
(20-50)~kpc, (50-100)~kpc and $>$100~kpc. Note in our measurements 
stars inside 7 kpc have been excluded (Sec.~\ref{sec:like}). The model 
distribution function is then fit to stars in each bin separately. 
However, in each case the three spatial parameters of the tracer 
distribution are fixed to their best-fit values obtained from tracers 
over the entire radial range, otherwise we would end up with extremely 
poor extrapolations based on the local density slope. All the other 
parameters, $M_{200}$, $c_{200}$ and $\beta$, are left as free parameters. 
The window function, Eqn.~\ref{eq:like}, is modified appropriately for each 
bin.

Fig.~\ref{fig:rbin} shows the measured $M_{200}$ as functions of the mean
radius of each bin, normalised by the halo virial radius ($R_{200}$).  The
value of $M_{200}$ varies significantly with the tracer radius. Thanks 
to the large number of stars (Table~\ref{tbl:num}), the errors are all quite 
small in each bin. For halos A, C and E, stars in the outermost ($r>$100~kpc) 
and innermost ($r<20$~kpc) bins give underestimates, while stars at 20$<r<$100~kpc 
give significant overestimates. Similarly, for halos B and D, stars at $r>$100~kpc 
give underestimates, whereas stars at smaller radii give overestimates. 

The velocity anisotropy of tracers, $\beta$, is a function of radius, whereas
the model distribution function assumes a single value of $\beta$.  To test
whether the radial average of $\beta$ may affect our estimates in the host halo
mass, we repeated the analysis of Fig.~\ref{fig:rbin} but fixed the value of
$\beta$ in each radial bin to either the best-fit value or the true value 
obtained from the whole population.  These measurements are almost identical to 
the measurements presented in Fig.~\ref{fig:rbin}, which confirms our result 
from Sec.~\ref{sec:cov} that the radial averaging of $\beta$ does not cause 
further bias in the other parameters.

One feature in Fig.~\ref{fig:rbin} is puzzling at first
glance: the best-fit halo masses obtained from the four radial bins
individually are all larger than the best-fit halo mass ($M_{200}=1.15$)
obtained using stars over the whole radial range. This seems odd, as we might
expect that the best-fit $M_{200}$ would be close to the average of the values
estimated from the four radial subsamples. The situation is not that
straightforward, however, because the likelihood surfaces from the subsamples 
are superimposed in two dimensional ($M_{200}$ and $c_{200}$) space when 
the full sample is used. Coupled with the strong correlation between the 
two parameters, the peak of the final likelihood surface is located around a 
region where the correlation lines from different subsamples intersect.

In real observations, there is often a maximum radius of tracers corresponding
to the instrumental flux limit. In the present literature this limit is
typically much smaller than the expected halo virial radius. Beyond this
maximum radius, extrapolations are required to fit the distributions of both
the dark matter and tracers.  We explore the implications of this directly in
Fig.~\ref{fig:rcut} by adopting several outer radial cuts ($r<r_\mathrm{cut}$)
and reporting the estimated halo mass as a function of the cut radius
normalised to the virial radius ($r_\mathrm{cut}/R_\mathrm{200}$). Unlike
Fig.~\ref{fig:rbin}, the three spatial parameters are treated as unknown and
left free to be constrained by the fit, in order to mimic real observations
where the density profiles of tracers is taken directly from the available
data.

The overall trends with $r_\mathrm{cut}$ in Fig.~\ref{fig:rcut} are very clear: 
the recovered halo mass is constant at large $r_\mathrm{cut}$, and turns up 
once $r_\mathrm{cut}$ becomes small (about 40 per cent of $R_\mathrm{200}$). 
We checked the best-fit values of the tracer spatial parameters in each case, 
and found they do not vary much with the radial cut as long as $r_{\rm cut}<0.4R_{200}$. 
This is because the break radius of tracer density profiles in our mock catalogues 
are smaller than $0.4R_\mathrm{200}$ for all the five halos, and so the extrapolation 
in tracer density is not severe. However, once $r_{\rm cut}$ reduces below $0.4R_{200}$, 
the outer slope becomes essentially unconstrained. We believe the turn-up behaviour 
is due to the changing dynamical state of tracers and the extrapolations required to 
know the underlying potential where there are no tracers.    

Previous constraints on the MW halo mass have been derived from tracers
roughly covering the range 0.1 to 0.4 $R_\mathrm{200}$
\citep[$R_{200}\sim$250~kpc;][]{2012MNRAS.425.2840D}. Our results suggest 
that the halo mass, $M_\mathrm{200}$, derived from the fitting distribution 
function of these tracers may be significantly biased even with respect to 
`asymptotic' results from the same method using all stars in the halo. 
Furthermore, instead of being a sharp cut, the radial selection functions 
of real surveys are often complicated, with non-trivial incompleteness as 
functions of radius and angular position. These selection effects may cause 
additional bias in the measured host halo mass.

\begin{figure}
\epsfig{figure=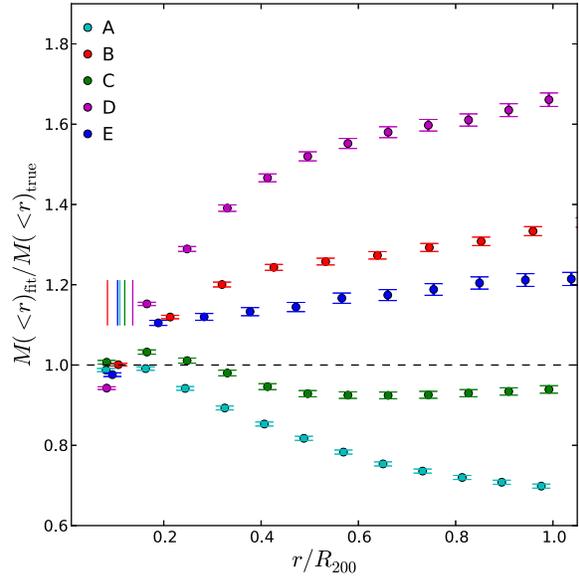,width=0.49\textwidth}%
\caption{The best-fit total mass within a fixed radius compared to the
  true mass within that radius, which is obtained from tracers inside 
  60~kpc and with both radial and tangential velocities. The 
  black dashed line marks equality between the measured and true mass. 
  Vertical lines mark the location of half-mass radii.}
\label{fig:massinsidet60}
\end{figure}

We have shown in Fig.~\ref{fig:massinside} that the total mass within
the half-mass radius of the stellar tracer population can be
constrained more precisely than the total mass of the halo. We now
test if this conclusion is robust to changes in the radial range
spanned by the stellar tracers. We repeated the measurements shown in
Fig.~\ref{fig:massinside} using only the stars within 60~kpc.  The
results are displayed in Fig.~\ref{fig:massinsidet60}, in the case
when both radial and tangential velocities are included in the
analysis. The total inferred mass within a fixed radius is strongly
biased if this radius is close to the virial radius, $R_\mathrm{200}$,
but, encouragingly, as the radius is decreased, the measured enclosed
mass becomes increasingly close to the true value. Our conclusion that
the mass enclosed within the half-mass radius can be constrained
reliably still holds even when stellar tracers inside only 60~kpc are
considered.

\section{Conclusions}
\label{sec:concl}

Several authors have measured parameters of the host halo of our MW, in
particular its total mass, by fitting specific forms of the distribution 
function to the observed distances and velocities of dynamical tracers such 
as old BHB and RR Lyrae stars, globular clusters and satellite galaxies
\citep{1999MNRAS.310..645W,2003A&A...397..899S, 2012MNRAS.424L..44D}. These 
models assume that the tracers are in dynamical equilibrium within the host 
potential. With the help of Jeans theorem, the distribution function of the 
tracers is further assumed to depend only on two integrals of motion, the 
binding energy, $E$, and the angular momentum, $L$.  In the case of a 
separable function of $E$ and $L$, the distribution function can be obtained 
through Eddington inversion of the tracer density profile.

In this paper we have extended earlier analytical forms of the MW halo
distribution function to the case of the NFW potential, which is of
most relevance to CDM-based models. We generalised the radial
distribution of tracers (halo stars) to a double power law, which is
suggested by recent observational results and simulations. We used a
maximum likelihood approach to fit this model distribution function to
a realistic mock stellar halo catalogue of distances and radial
velocities, constructed from the high resolution Aquarius $N$-body
simulations using the particle tagging technique of \cite{2010MNRAS.406..744C}.  
Our aim was to test the model performance and assumptions. We considered 
cases with and without additional tangential velocity data. Our conclusions 
are as follows:

\begin{itemize}

  \item The best-fit host halo virial masses and concentrations are biased from 
  the true values, with the level of bias varying from halo to halo. 
    
  \item Adding tangential velocity data substantially reduces this bias, but
    does not eliminate it. For example, for halo B the agreement between
    measured and true halo mass is very good (a 5\% overestimate) if tangential
    velocities are used, but for halo A, a 40\% underestimate persists even
    with this additional constraint. The inclusion of tangential velocities
    therefore is crucial for accurate measurements of both host halo and tracer
    properties, especially for the velocity anisotropies of the tracers.

  \item A strong negative correlation between the host halo mass and the halo
   concentration is found in our analysis.  
 
 \item The model gives a strongly biased measurement of the velocity
   anisotropies of stars.
   
 \item If tangential velocities are available, the correlation between $\beta$ 
 and all the other parameters are very weak. If only radial velocities are used, 
 $\beta$ is strongly correlated with other parameters and the bias in $\beta$
 will be propagated to these parameters. This is because when tangential 
 velocities are not available, we have to rely on the model functional form 
 to infer the unknown tangential component and hence $\beta$.
 
  \item Various sources contribute to the biased estimates of halo properties.
    Violation of the spherical assumption is relatively sub-dominant for the
    five Aquarius haloes. Violation of the dynamical equilibrium assumption,
    caused for example by streams, could affect the fits significantly,
    although we do not observe a systematic sign for the bias in $M_{200}$
    (that is, unrelaxed substructures cause underestimates in some cases and
    overestimates in others). 
 
 \item  When including tangential velocities, the systematic bias tends to
   happen along the correlation direction of $M_{200}$ and $c_{200}$ except for
   halo A.
   
  \item In contrast to the significantly biased measurements of $M_{200}$ 
  or $c_{200}$, the model gives good constraints on the total mass within 
  the half-mass radius of stellar tracers when including tangential 
  velocities.

\end{itemize}

The strong correlation between $M_{200}$ and $c_{200}$ arises
because changes in the corresponding tracer velocity distribution due 
to the increase of one of these parameters can be roughly compensated 
by the other. The correlation between $M_{200}$ and $c_{200}$ 
is not as strong for the radial velocity only case, which is probably 
overwhelmed by the strong correlation between $\beta$ and the other 
parameters, reflecting the fact that the dominant source of bias is 
the model dependent fit of the tangential component. If the model fails 
to properly reflect the true phase-space distribution of tracer objects, 
the best fit $\beta$ and other parameters will be strongly biased.

There are different combinations of halo parameters which give 
similar likelihood values along the correlation direction. Thus the 
model is vulnerable to perturbations (for example from dynamically hot 
structures). This can be seen from Fig.~\ref{fig:RTinfall}: the best 
fit $M_{200}$ and $c_{200}$ are offset in opposite directions with 
respect to their true parameter values. This is not the case for halo 
A, because the dominant source of bias for halo A is the 
deviation from the NFW model, and the error contour of $M_{200}$ and 
$c_{200}$ is not as elongated as the other halos. More detailed 
discussions about this halo will be given in a future study 
\citep{han2015b}.


It is, however, confusing to see that although the systematic bias tends to
happen along the correlation direction of $M_{200}$ and $c_{200}$, it is much
larger than the statistical errors. For example, we can see clearly in
Fig~\ref{fig:contourTA} that the best fit $M_{200}$ is about 15$\sigma$ away
from its true value. This is probably because the statistical error in our
analysis does not account for the correlations introduced by phase-space
structures or clumps. Our mock stellar halo catalogue contains a very large
number of stars. However, these individual stars are not completely independent
of each other. Structures such as coherent streams are highly correlated in
phase space and it is possible the true number of independent components is
much smaller than the total number of stars. To quantify the true number of
degrees of freedom by considering correlated phase structures is beyond the
scope of our current study. A more detailed discussion is given by
\cite{han2015a,han2015b}, in which we introduce a new method based on the
steady state assumption, independent of any other assumptions about the model
functional form.

Encouragingly, we found the mass within the half-mass radius of 
the tracer population to be relatively insensitive to the parameter 
correlations and can be constrained more robustly once tangential velocities 
are used. This is true even when only stars within about 60~kpc are available.
Similar correlations between model parameters and the robustness of the best 
constrained mass within a fixed radius have been reported and discussed in 
previous work \citep[e.g.][]{2014MNRAS.439.2678D,2014ApJ...794...59K,
2010MNRAS.406.1220W}, although these models are quite different from ours. 
For our model, the correlation could be closely related to the fact that 
there are relatively few stars outside $0.2R_\mathrm{200}$, beyond which 
the stellar radial profiles drop very quickly. Given the large number of 
dynamical tracers inside $0.2R_\mathrm{200}$, it is not surprising to find 
that the mass within this radius can be better constrained. In contrast, the 
total halo mass, $M_\mathrm{200}$, is dominated by mass in the outskirts of 
the halo and more tracers at large radii are required to have better 
constraints.

Further information needs to be incorporated into the model to weaken the 
correlation between mass and concentration. For example, including more tracers at 
large radii (perhaps from tidal streams) may help to weaken the correlation and 
improve the measurements of mass in the outer halo. Satellite galaxies and globular 
clusters in the outer parts of the halo with proper motion measurements could be useful. 
Having two populations of tracers at two different radial ranges could also be very 
helpful \citep[e.g.][]{2011ApJ...742...20W}. This would enable us to constrain the 
mass at two different half-mass radii and hence the entire mass profile can be fixed. 
More detailed investigations regarding the nature of correlation between $M_{200}$ 
and $c_{200}$ have been carried out by \cite{han2015a,han2015b}.

The correlations between velocity anisotropy, $\beta$, and all other parameters
are very weak when including tangential velocities. This is fortunate, because 
we know that the model can give systematically biased estimates of $\beta$ for 
stars; this particular bias is not propagated to the other parameters when 
including tangential velocities. However, if only radial velocities are available, 
the condition becomes very different. The correlation between $\beta$ and all the 
other parameters is strong. Combined with the biased measurements of $\beta$ in 
Fig.~\ref{fig:radialinfall}, this suggests that if proper motions are not available, 
it will be difficult to obtain robust constraints on $\beta$ and the bias may affect 
the fitting of the other parameters. Only by including tangential velocities can the 
correlation between $\beta$ and the other parameters be broken and $\beta$ be better 
constrained.

In addition to the correlation between halo mass and concentration,
relatively weak but still significant correlations exist between these
halo parameters and the three parameters describing the spatial
variation of tracer density. When including tangential velocities,
we found that a steeper inner slope gives a lower estimate of halo mass, 
while a steeper outer slope gives an higher estimate. If the true tracer 
density profile deviates from the double power-law form, the resulting 
bias will be propagated to the best-fit values of the halo parameters.

The model distribution function requires tracers to be in dynamical
equilibrium, with time-independent phase-space density. In reality, stars
stripped from satellite galaxies can have highly correlated orbits
that violate this assumption. We were able to test how well the
assumption holds for our mock halo stars. Perhaps surprisingly, we do
not find any systematic correlation of the recovered halo mass with
the infall redshift of tracer subsamples.  This suggests that the
dynamical state of halo tracers depends on other factors, such as
their orbits, and not only their infall time. Dynamical relaxation is
nevertheless a factor: excluding stars stripped from surviving
satellites improves the agreement between best-fit and true halo
masses in two cases (halos D and E). This cut eliminates dynamically
hot structures that can be identified by eye in these halos.

Beyond all these assumptions and uncertainties in the model itself, in
real observations the maximum observable radius of dynamical tracers
may be much smaller than the halo virial radius.  We found tracer
subsamples selected over different ranges of radius can give significantly
different estimates of the host halo mass, even if the three
parameters describing the density of tracers are fixed to be those
derived from the whole tracer population. An outer radius limit
results in biased measurements of $M_{200}$ if it significantly smaller 
than the virial radius. For example, the recovered halo masses of halos 
A, B, C and D converge for outer radius limits larger than $r\sim
0.4R_\mathrm{200}$ but give systematically larger masses for smaller
radial limits. For one halo, E, this overestimation occurs for limits
$r\lesssim 0.8R_\mathrm{200}$.  There are two reasons behind this
radial dependence: stars at different radii have significantly
different dynamical state and extrapolations to larger and smaller
radii become less accurate when only a limited radial range is
sampled.

Real surveys have complex selection functions for stars, which depend on
both radial distance and angular position. Particular classes of tracers may be
very sparsely sampled. The observed parallax, radial and tangential velocities
of halo stars include observational uncertainties which depend strongly on
distance. Although a large sample of tracers with exact coordinate and velocity
information from our mock stellar halo catalogue have enabled us to investigate
the model performance, it will be important to consider realistic observational
errors and sample selection effects in future studies aimed at forecasting the
performance of real surveys.

We conclude that methods to estimate the mass of the Milky Way halo
using the kind of distribution functions we have investigated here
need to be used with extreme caution. This is particularly true when
estimating the total virial mass.  Restricting the estimate to the
mass interior to $r\sim 0.2R_\mathrm{200}$ is considerably more
reliable. In any case, mock catalogues like those we have analysed 
here and made publicly available in \cite{2015MNRAS.446.2274L}
are required to assess the reliability of any particular mass estimation
method. 

\section*{Acknowledgements}

We sincerely thank the anonymous referee for his/her useful 
suggestions that helped to improve the paper significantly.
This work was supported by the Euopean Research Council [grant number
GA 267291] COSMIWAY and Science and Technology Facilities Council
Durham Consolidated Grant. WW and APC acknowledges the Durham 
Junior Research Fellowship. The simulations for the Aquarius project 
were carried out at the Leibniz Computing Centre, Garching, Germany, 
at the Computing Centre of the Max-Planck-Society in Garching, at the 
Institute for Computational Cosmology in Durham, and on the STELLA 
supercomputer of the LOFAR experiment at the University of Groningen. 
This work used the DiRAC Data Centric system at Durham University, 
operated by the Institute for Computational Cosmology on behalf of 
the STFC DiRAC HPC Facility (www.dirac.ac.uk).  This equipment was 
funded by BIS National E-infrastructure capital grant ST/K00042X/1, 
STFC capital grant ST/H008519/1, and STFC DiRAC Operations grant 
ST/K003267/1 and Durham University. DiRAC is part of the National 
E-Infrastructure. WW is grateful for useful discussions with 
Yanchuan Cai, Yipeng Jing, Till Sawala and John Helly. 

\bibliography{master}


\appendix

\section{Origin of the bias in $\beta$}
\label{sec:betabias}

\begin{figure*}
\epsfig{figure=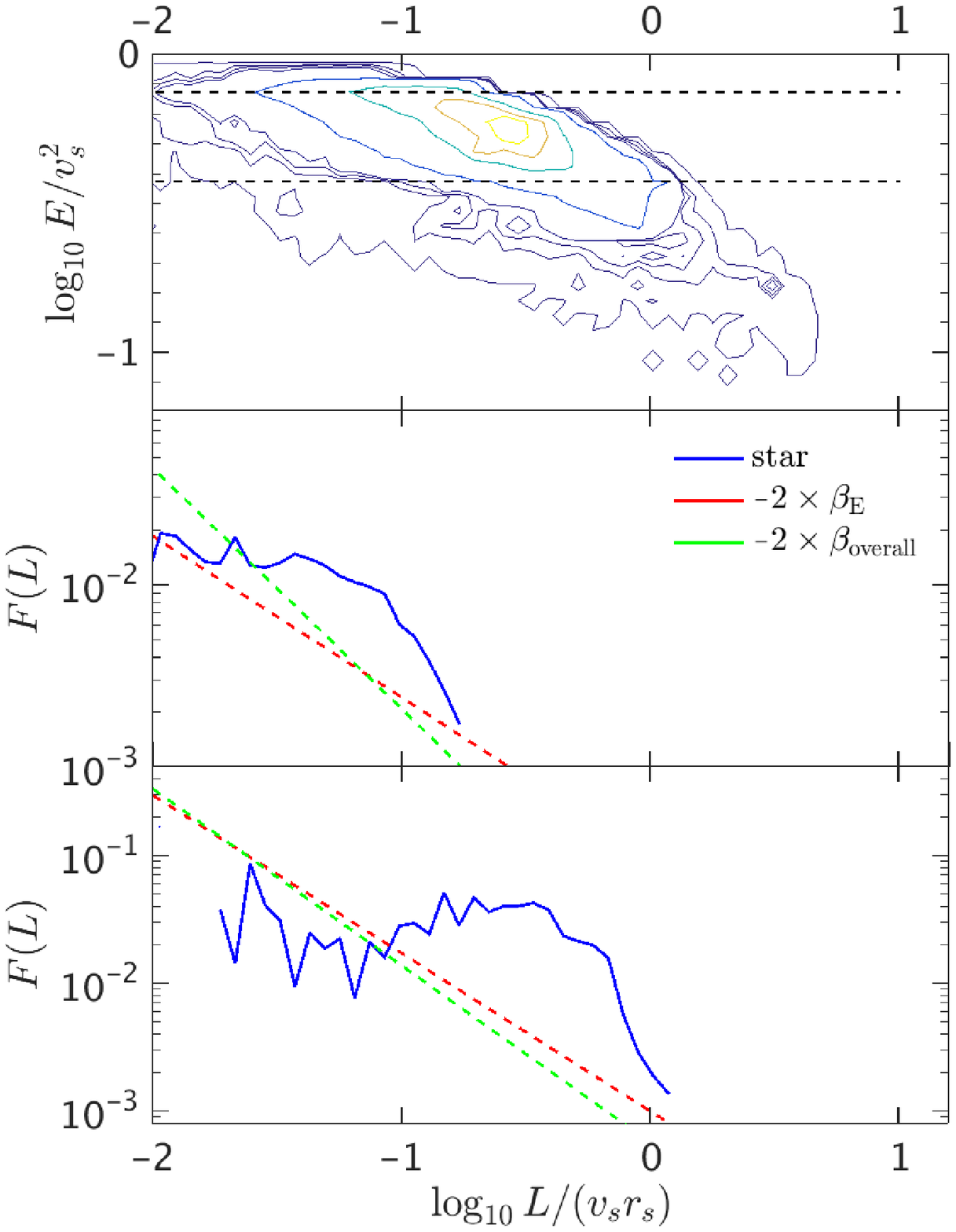,width=0.49\textwidth, height=10cm}%
\epsfig{figure=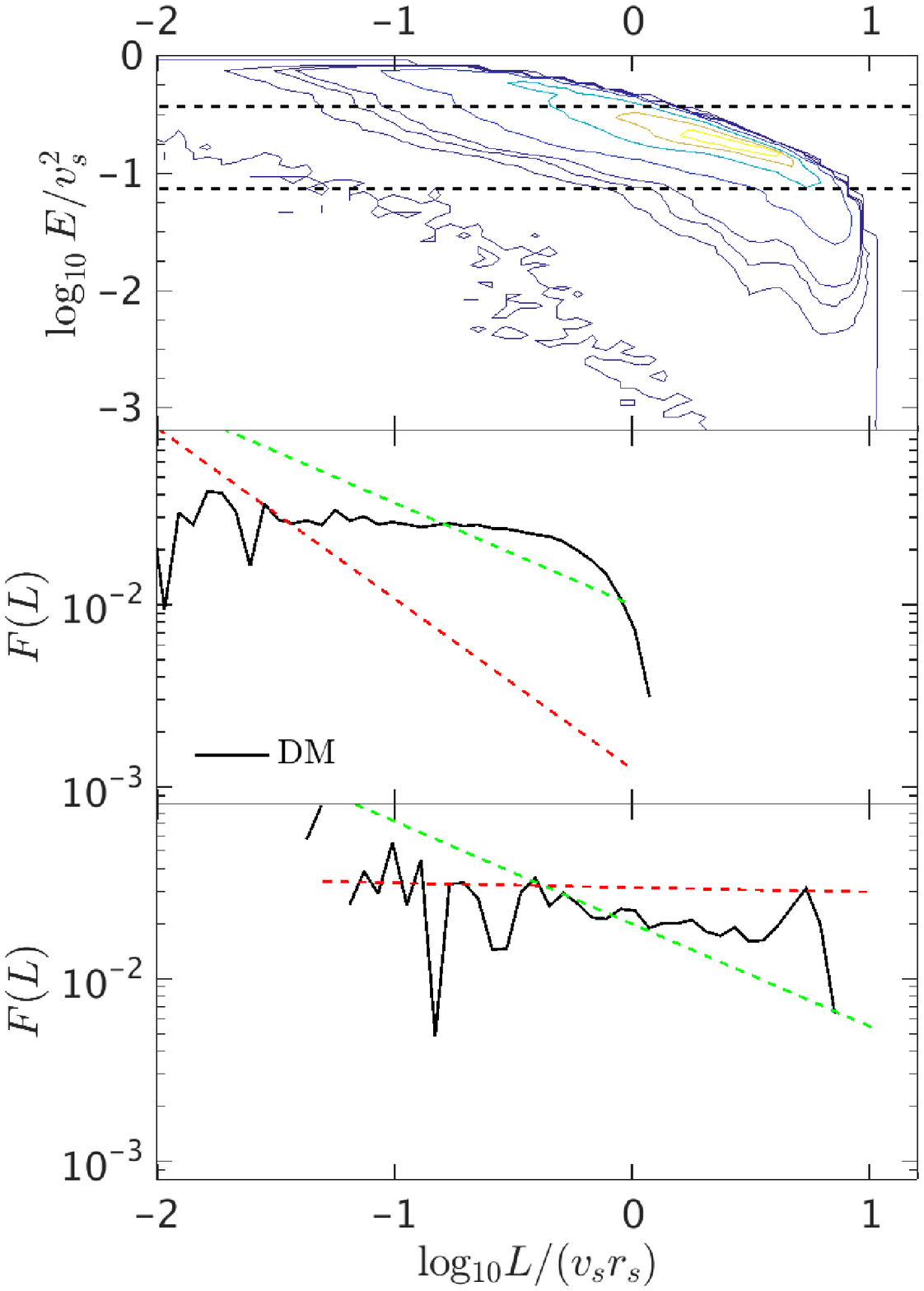,width=0.49\textwidth, height=10cm}
\caption{phase-space distributions of mock stars in halo A (left) and of 
dark matter in the same halo (right). The top panels in either plots show 2D 
distributions in the plane of binding energy, $E$, and angular momentum, $L$. 
$E$ and $L$ have been scaled to be dimensionless. The middle and bottom panels
show 1D angular momentum distributions of stars (likewise DM) at two fixed
values of $E$, indicated by the horizontal black dashed curves in the top
panels. Red and green dashed lines are power-law distributions with arbitrary
normalizations, predicted from the velocity anisotropy of all stars (or DM) and
stars (or DM) in each binding energy range.}
\label{fig:histL}
\end{figure*}

\subsection{Origin of the bias}

In Fig.~\ref{fig:RTinfall} we showed that there is a systematic bias
between the best-fit and true value of $\beta$. We now show in the
left panels of Fig.~\ref{fig:histL} the phase-space distribution of
stars in halo A (top panel, binding energy, $E$, versus angular
momentum, $L$) and the one dimensional angular momentum distribution
at two fixed values of $E$ (the second and third panels from the top,
respectively) indicated by the horizontal dashed lines in the top
panel.\footnote{Note the quantity we are plotting here is
  $F(L)=F(L|E)$. To obtain this distribution empirically one has to 
properly account for the density of state in ($E$,$L$) 
space~\citep[see][for more details]{2008MNRAS.388..815W}.}  $E$ and $L$ have
been normalised so that they are dimensionless (see
Eqn.~\ref{eq:dimless}). We only show results based on halo A; for the
other halos our conclusions are the same.

In the top panel we see that, for a fixed value of $E$, there is an upper limit
to $L$ which increases with decreasing $E$. This is the maximum allowed value
of angular momentum, corresponding to circular orbits with zero radial velocity
at fixed $E$. In the next two panels, we show the angular momentum distributions 
at different values of $E$ have similar features. They are both flat at small values 
of $L$ and drop quickly when $L$ approaches its upper limit.  For comparison, we also 
plot two lines of the form $F(E,L) \propto L^{-2\beta}$, where $\beta$ is the velocity 
anisotropy obtained from stars in the energy slice, $\beta_\mathrm{E}$ (red dashed 
lines), or the full sample, $\beta_\mathrm{overall}$ (green dashed lines). 
$\beta_\mathrm{overall}$ is the same in both panels and is simply the true value 
of $\beta$ from Table~\ref{tbl:parameters} (0.66). The values of $\beta_\mathrm{E}$ 
are 0.442 and 0.618 for the middle and bottom panels respectively. Neither of the 
two could give a satisfactory description of the true distribution (blue curve). 
This implies that the physical interpretation of the power-law index in our 
distribution function as $-2\beta$ is inaccurate, and this is the origin for the 
systematic bias of $\beta$. The true distribution function must be more complex. 

We also notice that the best-fit value of $\beta$ for halo A is $0.458$
(Table~\ref{tbl:parameters}), which predicts a power-law slope  
shallower than the green dashed lines and is also shallower than the 
red dashed line in the bottom panel, but it is still a poor match 
to the distributions (blue curve) in Fig.~\ref{fig:histL}. If we fix the 
power-law slope in the model according to the true anisotropy of the full 
sample, this results in better agreement with tangential velocity 
distribution but a much poorer agreement with the
radial velocity distribution. Afterall, our maximum likelihood approach is
designed to fit the velocity and spatial distributions of stars, not the
distributions of binding energy or angular momentum.

\subsection{Why stars are more radially biased than dark matter?}

The $\beta$ profile of dark matter particles have been studied in earlier
works. For example, \cite{2008MNRAS.388..815W, 2009MNRAS.399..812W} looked at
the distribution functions of dark matter particles in halos of mass $10^{14}$
to $10^{15} M_\odot$. Although the details of their modelling and the mass range
of halos are different from ours, their model distribution function can
recover well the true $\beta$ of dark matter particles in their simulation.
We therefore examine the angular momentum distribution of dark
matter particles in our simulations in the three right panels of
Fig.~\ref{fig:histL}. 


In the top panel, we see dark matter particles can extend to much lower binding energy 
than stars. Black curves in the middle and bottom panels show the angular momentum distribution 
at two fixed values of $E$. We again plot two lines of the form $F(E,L) \propto L^{-2\beta}$. 
Red and green lines are predicted from the velocity anisotropy of dark matter particles in 
the energy slice or from the the full sample respectively. At $E\sim 10^{-0.4}v_s^2$, 
the agreement between the red dashed lines and the shape of the $L$ distributions is 
quite poor, whereas at a lower binding energy ($E\sim 10^{-1.1}v_s^2$), we see a better
agreement. We have looked at many different choices of $E$ in this regard, and
found that for less bound dark matter particles, their velocity anisotropy
correctly predicts the power-law slope of their $L$ distribution. This means the 
model distribution function describes better systems of less bound dark matter 
particles. However, for dark matter particles that are more tightly bound, the 
velocity anisotropy is not as well correlated with the power-law slope of the $L$ 
distribution. This is the same as the stellar case, although the discrepancy for dark 
matter particles is smaller. 

Stars in the stellar halo are clearly a biased population of tracers with respect 
to dark matter particles in the simulation. Their orbits are more radial 
(Fig.~\ref{fig:beta}) corresponding to a higher $\beta$. However, the difference 
in $\beta$ is not only because stars are more dynamically bound than dark matter 
particles: we have explicitly checked that, for a given fraction of the most 
bound dark matter particles, orbits are still more tangentially biased than stars 
with the same range of binding energy. The fact that stars are more radially 
biased than dark matter particles thus has more fundamental physical origin. 
First of all, in our model halo stars are all accreted from subhalos, while
dark matter particles are added to the main halo by both clumpy and smooth
accretion. We have calculated the velocity anisotropy of dark matter accreted
from subhalos only, and found these particles are more radially biased than
all the dark matter particles as a whole. This is probably because the clumps
in which these particles are accreted (i.e. subhalos) have more radially
biased orbits. Furthermore, halo stars in our analysis are tags placed on the
most bound dark matter particles in progenitor subhalos, which have then been
stripped and mixed into the main halo.  \cite{2015MNRAS.446.2274L} have found
the halo stars are dominated by contributions from a few massive satellites. As
the most bound parts of these satellites have been stripped into the halo, the
satellites are more likely to have been on highly radial orbits, imparting a
radial bias to halo stars. In contrast, dark matter particles enter the main
halos in our simulations through quite different mechanisms, with both clumpy
and smooth accretion \citep{2011MNRAS.413.1373W}.

\section{Uncertainties in modelling the halo boundaries}
\label{sec:fini}

For all the analysis in the main text, we have been assuming the spatial 
extent of both NFW halos and tracers are infinite  
($r_\mathrm{max,h}=\infty$ and $r_\mathrm{max,t}=\infty$). It is, however, 
necessary to investigate whether the different choices of halo and tracer
boundaries could affect our measured halo properties. 

As we have mentioned in Sec.~\ref{sec:nfw}, in principle tracer boundaries 
($r_\mathrm{max,t}$) can be different from halo boundaries. Here for simplicity 
we assume $r_\mathrm{max,h}=r_\mathrm{max,t}$. We tried four different choices 
of halo boundaries ($r_\mathrm{max,h}$), ranging from twice to five times the 
halo virial radius. We avoid using boundaries at exactly the halo virial radius 
because our mock halo stars can extend beyond $R_{200}$, while the mass 
distribution in the FoF group distribute continuously and extend further than 
$R_{200}$. A sharp cut at $R_{200}$ is thus not realistic. 

\begin{figure*}
\epsfig{figure=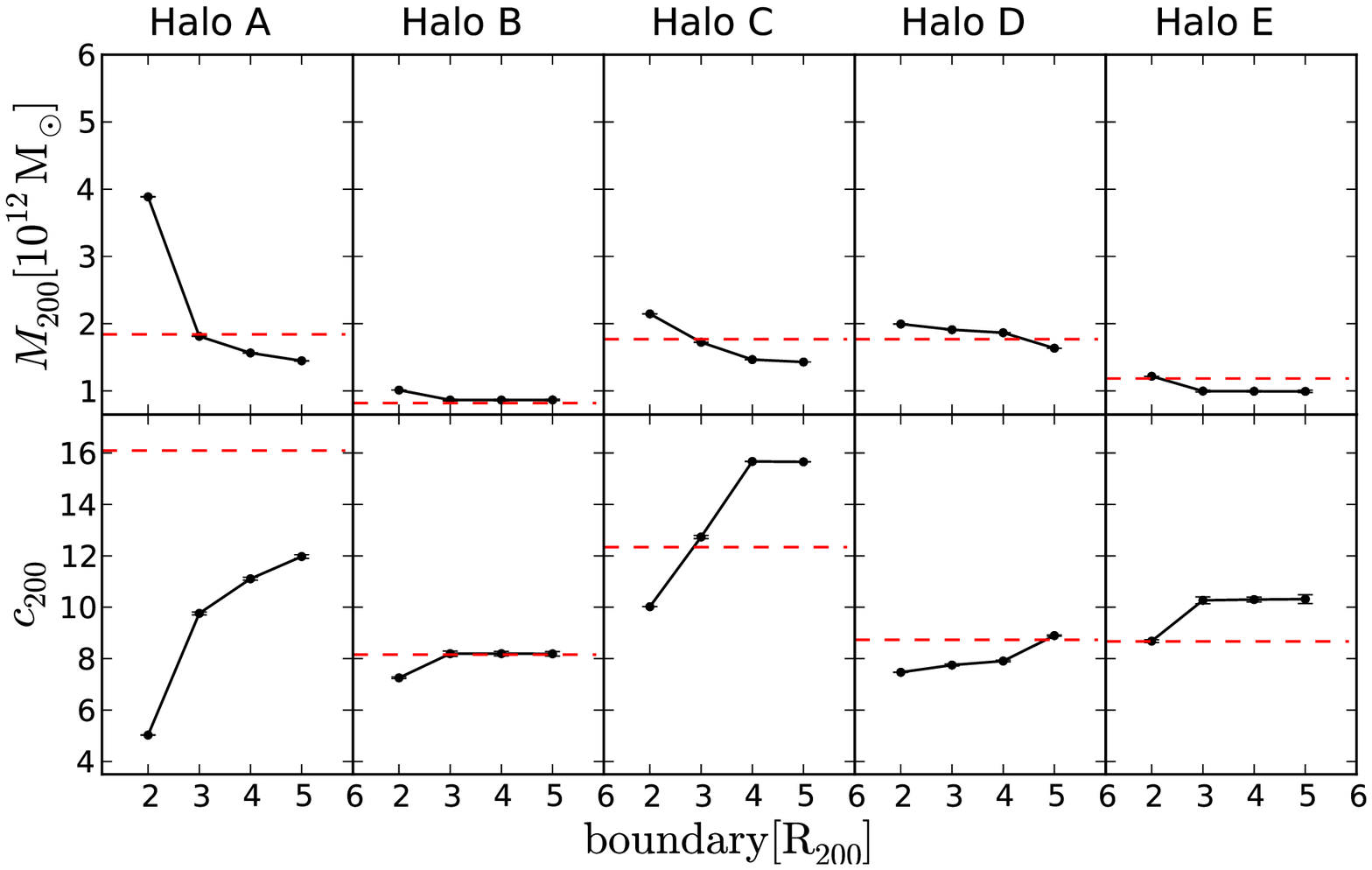,width=0.9\textwidth}
\caption{The measured host halo mass (top row) and concentration (bottom row) 
as functions of the halo boundaries in the model.}
\label{fig:fini}
\end{figure*}

The best-fit host halo masses and concentrations as functions of halo boundaries 
are presented in Fig.~\ref{fig:fini}. The velocity anisotropy $\beta$ almost does 
not change with the different choices of halo boundaries, and thus we do not show 
them. Dashed red lines are true values of halo masses and concentrations.

The measured halo masses increase with the decrease in halo boundaries, and the 
halo concentrations decrease accordingly, reflecting again the strong correlation
between the two parameters. The choice of halo boundary that gives the best match 
between measured and true halo mass varies from halo to halo. For halo B, the best 
fit halo masses and concentrations almost do not change with the choice of halo 
boundaries when $r_\mathrm{max,h} \geq 3 R_\mathrm{200}$ and agree well with the 
true values. At $r_\mathrm{max,h} = 2 R_\mathrm{200}$, the measured host halo mass 
gets significantly larger, indicating for halo B finite halo boundaries do not help 
to improve the fitting. For halo C, $r_\mathrm{max,h} = 2 R_\mathrm{200}$ gives a 
very good match between the best-fit and true halo mass and concentration, 
demonstrating a finite halo boundary works better than infinite boundaries at least 
for halo C. 

The estimated halo mass of halo A is closest to the true value when halo boundary is 
chosen at twice the virial radius. However, the estimated halo concentration at that 
boundary deviates significantly from the true concentration, suggesting the discrepancy 
between best-fit and true host halo mass of halo A could not be dominated by how 
boundaries are modeled. For halo D and E, we know already because of the existence of 
unrelaxed dynamical structures, the host halo masses are significantly underestimated, 
and thus the best match between measured and true halo parameters at twice (halo E) and 
four times (halo D) the virial radius demonstrates the entangling of different model 
uncertainties, which canceled with each other to give a good prediction. 

\section{Dark matter particles as tracers}
\label{App:D}

\begin{table*}
\caption{best-fit parameters from the phase-space distribution of dark matter particles}
\begin{center}
\begin{tabular}{lrrrrr}\hline\hline
           & \multicolumn{1}{c}{A} & \multicolumn{1}{c}{B} & \multicolumn{1}{c}{C} & \multicolumn{1}{c}{D} & \multicolumn{1}{c}{E}\\ \hline
$M_{200} [\mathrm{10^{12} M_\odot}]$ & 2.811 $\pm$ 0.024 &  1.227 $\pm$ 0.012 & 2.868 $\pm$ 0.024 & 2.727 $\pm$ 0.029 & 1.776 $\pm$ 0.015\\
$c_{200}$ & 4.458 $\pm$ 0.098 & 4.845 $\pm$ 0.105 & 6.031 $\pm$ 0.106 & 3.903 $\pm$ 0.104 & 5.296 $\pm$ 0.108 \\
$r_s[\mathrm{kpc}]$ & 63.499 $\pm$ 1.116 & 44.315 $\pm$ 0.773 & 47.257 $\pm$ 0.683 & 71.799 $\pm$ 1.516 & 45.867 $\pm$ 0.757 \\
$\log_{10}\rho_s[\mathrm{M_\odot/kpc^3}]$ & 5.997 $\pm$ 0.021 & 6.078 $\pm$ 0.021 & 6.297 $\pm$ 0.018 & 5.868 $\pm$ 0.025 & 6.166 $\pm$ 0.020 \\
$R_{200}[\mathrm{kpc}]$ & 283.078 $\pm$ 7.965 & 214.715 $\pm$ 5.956 & 285.003 $\pm$ 6.482 & 280.225 $\pm$ 9.529 & 242.901 $\pm$ 6.357 \\
$\beta$ & 0.266 $\pm$ 0.004 & 0.159 $\pm$ 0.005 & 0.259 $\pm$ 0.005 & 0.217 $\pm$ 0.006 & 0.090 $\pm$ 0.005 \\
\hline
\label{tbl:DM}
\end{tabular}
\end{center}
\end{table*}

In Sec.~\ref{sec:betabias} we show the velocity anisotropies of dark matter
particles agree better with the distribution function model than those of
stars. Furthermore, dark matter particles are more radially extended than stars 
and might probe better the underlying potential in outskirts. Thus we ask whether 
better constraints on the halo properties can be achieved by using dark matter 
particles as tracers. Obviously, it is not possible to directly observe the 
dynamics of dark matter, but asking this question helps to deepen our understanding 
of the model. The answer is, unfortunately, {\it no}. By using  
dark matter particles as tracers we end up with significant overestimates 
of the host halo mass, at least for our five Milky Way analog halos. These measurements 
are shown in Table~\ref{tbl:DM}, where we have used a randomly selected subsample
of all dark matter particles in the halo FoF group (one particle out of every
5,000 in the simulation). We have explicitly checked that this conclusion does
not change if we randomly select different subsamples of dark matter, remove
dark matter particles in substructures or restrict them to be inside the halo
virial radius. Both radial and tangential velocities have been used in this
analysis.   

\begin{figure}
\epsfig{figure=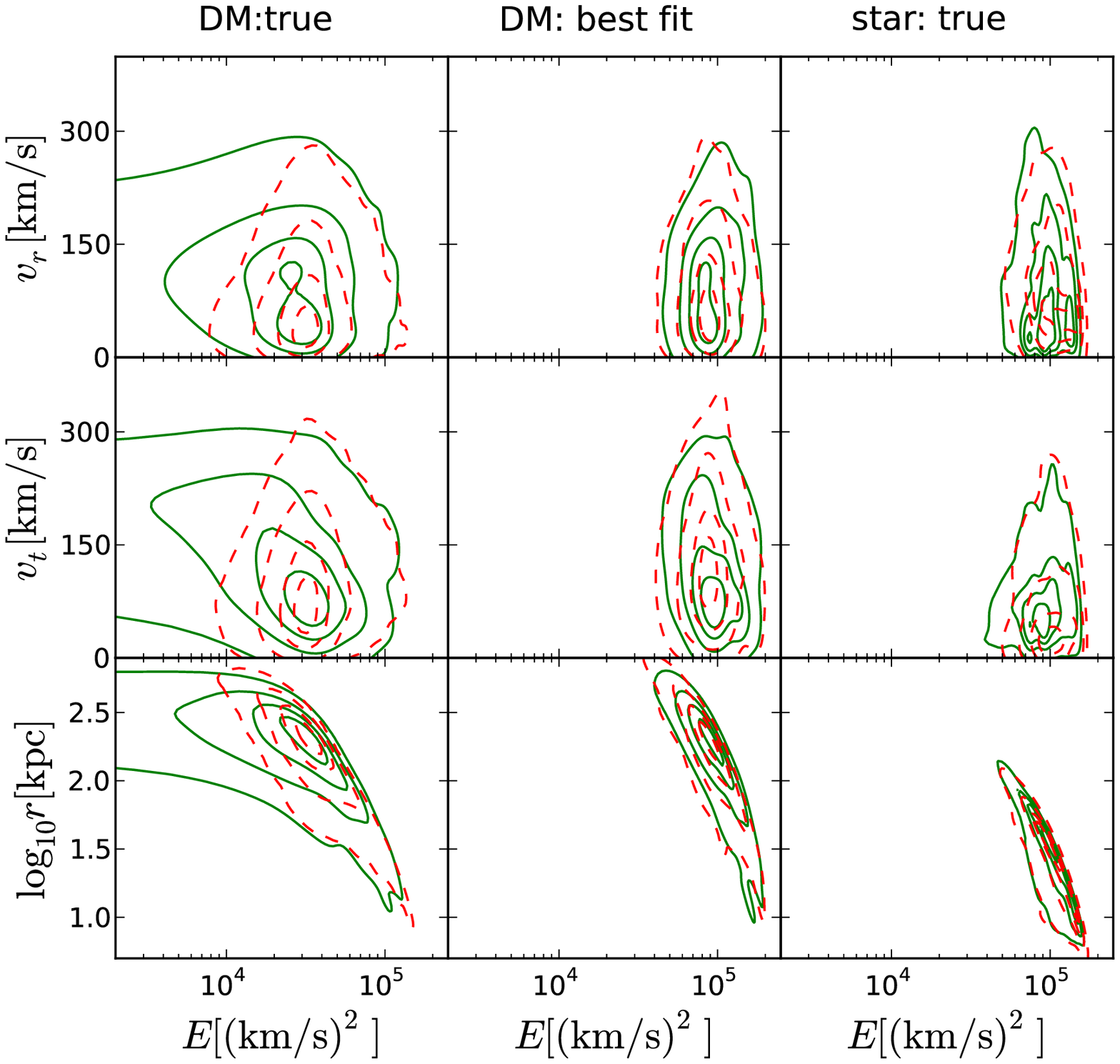,width=0.49\textwidth}
\caption{The phase-space density of dark matter particles or stars in halo 
A (green solid) and model predictions (red dashed). The contours mark the 10th, 30th, 
60th and 90th percentiles of the 2D density distribution in  parameter plane. We present 
contour plots of binding energy, $E$, versus radius, $r$, radial velocity, $v_r$, and 
tangential velocity, $v_t$. For simplicity we only use the magnitudes of $v_r$ and $v_t$, 
so all quantities are positive. In deducing the binding energy, we use the analytical 
NFW potential model. Green contours in the left and middle columns are based on 
dark matter particles in the simulation, while in the right column we plot 
contours for stars. For the left, middle and right columns, true halo parameters,
dynamical best-fit halo parameters from dark matter particles and true halo parameters 
are adopted in the potential model respectively. The lines are iso-density contours 
that contain the 10,30,60 and 90\% densest cells. }
\label{fig:contourDM}
\end{figure}

To explore the reasons behind this, we present in Fig.~\ref{fig:contourDM}
phase-space contour plots for dark matter particles and stars in the simulation
and compare these with realizations drawn directly from the model distribution
function. We only show plots based on halo A; for the other halos the
conclusion is the same.  Distributions of binding energy versus radial
velocity, $v_r$, tangential velocity, $v_t$, and radius, $r$, have been plotted
separately, so that we are able to see how well the model prediction agrees
with the true distribution of $v_r$, $v_t$ and $r$ for dark matter particles in
the simulation. 

It is very clear to see that, with true halo parameters, the model predictions
deviate significantly from the empirical distribution of $r$, $v_r$ and $v_t$ at 
low binding energy (left column). On the other hand, the best fit model agrees 
much better with the data (middle column). This improved agreement is caused by 
an overestimate of the host halo mass, leading to a deeper potential and increased 
binding energy for tracer particles. As a result, the sample becomes more dynamically 
bound and agrees better with the model.

Our conclusion is therefore that, although the model distribution function
gives a better approximation in the velocity anisotropy of dark
matter particles, the predicated phase-space distribution at the low binding 
energy end is very poor. By construction, the distribution function only describes 
closed systems. This not only requires that the tracer population should be bound 
and truncated, but also results in a sharp cut-off in binding energy that is a 
poor description of particle distribution in the low energy end. Compared with 
all dark matter particles, the star particles in our simulation are more 
dynamically bound and centrally concentrated (see the right column) and 
are thus not sensitive to the low binding energy tail of the dark
matter distribution. Using stars rather than dark matter particles
therefore insulates the fit from deficiencies of the distribution function
model at the low binding energy tail.

\end{document}